\documentclass[iop,revtex4]{emulateapj}
\slugcomment{Accepted to ApJ: 13 March 2018}

\usepackage[backref,breaklinks,colorlinks,citecolor=blue]{hyperref}
\usepackage[all]{hypcap}
\usepackage{graphicx} 
\usepackage{mathrsfs} 
\usepackage{natbib}
\usepackage{amsmath}
\usepackage{float}
\usepackage{amssymb}
\usepackage{color}
\usepackage{captcont}

\newcommand{\as}{$^{\prime\prime}$}

\begin{document}

\title{ALMA Survey of Lupus Protoplanetary Disks II: Gas Disk Radii}

\author{M. Ansdell$^{1,2}$, J. P. Williams$^{1}$, L. Trapman$^{3}$, S. E. van Terwisga$^{3}$, S. Facchini$^{4}$, C.F. Manara$^{5}$, N. van der Marel$^{1, 6}$, A. Miotello$^{5}$, M. Tazzari$^{7}$, M. Hogerheijde$^{3, 8}$, G. Guidi$^{9}$, L. Testi$^{5,9}$, E. F. van Dishoeck$^{3,4}$}
 
\affil{$^1$Institute for Astronomy, University of Hawai`i at M\={a}noa, Honolulu, HI 96822, USA}
\affil{$^2$Department of Astronomy, University of California, Berkeley, CA 94720, USA}
\affil{$^3$Leiden Observatory, Leiden University, PO Box 9513, 2300 RA Leiden, The Netherlands}
\affil{$^4$Max-Plank-Institut f\"{u}r Extraterrestrische Physik, Giessenbachstra\ss e 1, D-85748 Garching, Germany}
\affil{$^5$European Southern Observatory, Karl-Schwarzschild-Str. 2, D-85748 Garching bei M\"{u}nchen, Germany}
\affil{$^6$Herzberg Astronomy \& Astrophysics Programs, NRC of Canada, 5017 West Saanich Road, Victoria, BC V9E 2E7, Canada}
\affil{$^7$Institute of Astronomy, University of Cambridge, Madingley Road, CB3 0HA, Cambridge, UK}
\affil{$^8$Anton Pannekoek Institute for Astronomy, University of Amsterdam, Science Park 904, 1098 XH Amsterdam, Netherlands}
\affil{$^9$INAF-Osservatorio Astrofisico di Arcetri, Largo E. Fermi 5, I-50125 Firenze, Italy}

%========================= ABSTRACT =========================

\begin{abstract}

We present ALMA Band~6 observations of a complete sample of protoplanetary disks in the young ($\sim$1--3~Myr) Lupus star-forming region, covering the 1.33~mm continuum and the $^{12}$CO, $^{13}$CO, and C$^{18}$O $J=2$--1 lines. The spatial resolution is $\sim0\farcs25$ with a medium 3$\sigma$ continuum sensitivity of 0.30~mJy, corresponding to $M_{\rm dust}\sim0.2~M_{\oplus}$. We apply ``Keplerian masking" to enhance the signal-to-noise ratios of our $^{12}$CO zero-moment maps, enabling measurements of gas disk radii for 22 Lupus disks; we find that gas disks are universally larger than mm dust disks by a factor of two on average, likely due to a combination of the optically thick gas emission as well as the growth and inward drift of the dust. Using the gas disk radii, we calculate the dimensionless viscosity parameter, $\alpha_{\rm visc}$, finding a broad distribution and no correlations with other disk or stellar parameters, suggesting that viscous processes have not yet established quasi-steady states in Lupus disks. By combining our 1.33~mm continuum fluxes with our previous 890~$\mu$m continuum observations, we also calculate the mm spectral index, $\alpha_{\rm mm}$, for 70 Lupus disks; we find an anti-correlation between $\alpha_{\rm mm}$ and mm flux for low-mass disks ($M_{\rm dust} \lesssim 5$), followed by a flattening as disks approach $\alpha_{\rm mm} \approx 2$, which could indicate faster grain growth in higher-mass disks, but may also reflect their larger optically thick components. In sum, this work demonstrates the continuous stream of new insights into disk evolution and planet formation that can be gleaned from unbiased ALMA disk surveys.

\end{abstract}

\maketitle

%======================= INTRODUCTION ===================================

\section{INTRODUCTION}
\label{sec-intro}

Thousands of exoplanet systems have now been detected and characterized, yet exactly how these planets formed remains unclear due to our still incomplete understanding of the structure and evolution of the preceding protoplanetary disks \cite[e.g.,][]{2016JGRE..121.1962M}. Early infrared (IR) surveys of nearby star-forming regions, which probed unresolved and optically thick inner disk emission, revealed that protoplanetary disks disperse quickly, typically within $\sim$5--10~Myr \cite[e.g.,][]{2007ApJ...662.1067H}. The specifics of this dispersal, however, are still needed to understand how disks evolve into planetary systems. The Atacama Large Millimeter/Sub-Millimeter Array (ALMA) is now enabling high-resolution and high-sensitivity sub-mm/mm observations of optically thin disk emission in both the continuum and line. The combination of these ALMA observations with other state-of-the-art datasets, in particular those from facilities like VLT/X-Shooter for constraining host star properties, is providing the needed insights into disk evolutionary processes \citep{2016ApJ...828...46A, 2016A&A...591L...3M, 2016ApJ...831..125P, 2017MNRAS.472.4700L, 2017ApJ...847...31M, 2017MNRAS.468.1631R}.

Moreover, large-scale ALMA surveys of nearby star-forming regions with ages spanning the disk lifetime ($\sim$1--10~Myr) are providing quantitative characterizations of disk dispersal and revealing statistical properties that can be linked to exoplanet trends. Combining the recent ALMA surveys of the protoplanetary disk populations in the young ($\sim$1--3~Myr) Lupus \citep{2016ApJ...828...46A} and Chamaeleon I \citep{2016ApJ...831..125P} regions, with those of the intermediate-aged ($\sim$3--5 Myr) $\sigma$~Orionis cluster \citep{2017AJ....153..240A} and the evolved ($\sim$5--10~Myr) Upper Sco association \citep{2016ApJ...827..142B}, reveal a clear decline in disk dust mass ($M_{\rm dust}$) with age \cite[see Figure 8 in][]{2017AJ....153..240A}. Even at just a few Myr of age, only $\sim$25\% of disks have sufficient reservoirs of dust to form giant planet cores ($M_{\rm dust}\gtrsim10~M_\oplus$), in line with the rarity of giant planets seen in the exoplanet population \cite[e.g.,][]{2012Natur.481..167C,2013ApJ...766...81F,2014ApJ...781...28M,2015ApJS..216....7B,2016MNRAS.457.2877G}. Alternatively, large amounts of solids could be rapidly locked into larger bodies, such as pebbles and planetesimal, which go undetected in these sub-mm/mm surveys that can only probe dust grains up to roughly cm sizes; this scenario is more consistent with evidence from our own solar system, which points to the formation of mm- to cm-sized chondrules \cite[e.g.,][]{2008ApJ...675L.121C} and even the differentiation of asteroids \cite[e.g.,][]{2002Natur.418..952K} within just a few Myr. Yet another possibility is that unconstrained amounts of dust are being hidden in the optically thick inner disk regions due to the growth and inward radial drift of the dust \citep{1977Ap&SS..51..153W}. Disentangling these scenarios will be critical to understanding the timescales of disk evolution and planet formation.

Another important property is the disk size, which is a fundamental input into planet formation models that can also be used to distinguish between different disk evolutionary pathways. Disks are traditionally thought to evolve through viscous accretion \citep{1974MNRAS.168..603L}, which predicts that gaseous disks spread outward with age due to the re-distribution of angular momentum to counter the accretion of disk material onto the star. Indeed, \cite{2017A&A...606A..88T} found that Lupus disks tend to be larger and less massive than slightly younger Taurus and $\rho$~Ophiuchus disks, which they tentatively attribute to viscous evolution. The growth and inward radial drift of solids \citep{2014ApJ...780..153B}, potentially in combination with optical depth effects \citep{1998A&A...339..467G, 2017A&A...605A..16F}, can also make the dust disk appear smaller than the gas disk at sub-mm/mm wavelengths, as seen for several individual disks  \citep{2007A&A...469..213I, 2009A&A...501..269P, 2012ApJ...744..162A, 2013A&A...557A.133D, 2016ApJ...832..110C}. If magnetohydrodynamic (MHD) winds play an important role in disk evolution \cite[e.g.,][]{2016ApJ...818..152B}, then they may suppress the viscous spreading of disks by removing angular momentum (rather than redistributing it outward), which could help to explain the surprisingly small dust disk radii ($\lesssim$20~au) seen in the ALMA surveys of Lupus, $\rho$~Ophiuchus, and Upper Sco \citep{2017A&A...606A..88T, 2017arXiv171103974C, 2017arXiv171104045B}.

In the first paper of this series (\citealt{2016ApJ...828...46A}; hereafter Paper~I), we used ALMA to observe a near-complete sample of protoplanetary disks in the Lupus star-forming region in the 890~$\mu$m (Band~7) continuum as well as the $^{13}$CO and C$^{18}$O $J=3$--2 isotopologue lines. The focus of Paper~I was to constrain both dust and gas masses for a large, unbiased population of protoplanetary disks within a single star-forming region, allowing us to perform statistical studies related to disk mass. In this work, we present new ALMA observations of the complete sample of Lupus protoplanetary disks in the 1.33~mm (Band~6) continuum as well as the $^{12}$CO, $^{13}$CO, and C$^{18}$O $J=2$--1 lines. These new data, in particular with the addition of $^{12}$CO, now allow us to add gas disk size to our statistical studies. Moreover, when also considering our previous Band~7 data, we can study processes such as viscous evolution as well as dust grain growth and radial drift. 

We describe our sample in Section~2, present our ALMA observations in Section~3, and give the measured continuum and line fluxes in Section~4. In Section~5, we measure disk radii and disk masses as well as identify individual objects of interest. Our findings are discussed in the context of disk evolution and planet formation in Section~6, then our work is summarized in Section~7.

 %======================= LUPUS SAMPLE  ==================================

\section{LUPUS SAMPLE}
\label{sec-sample}

 \capstartfalse
\begin{figure}
\begin{centering}
\includegraphics[width=8.5cm]{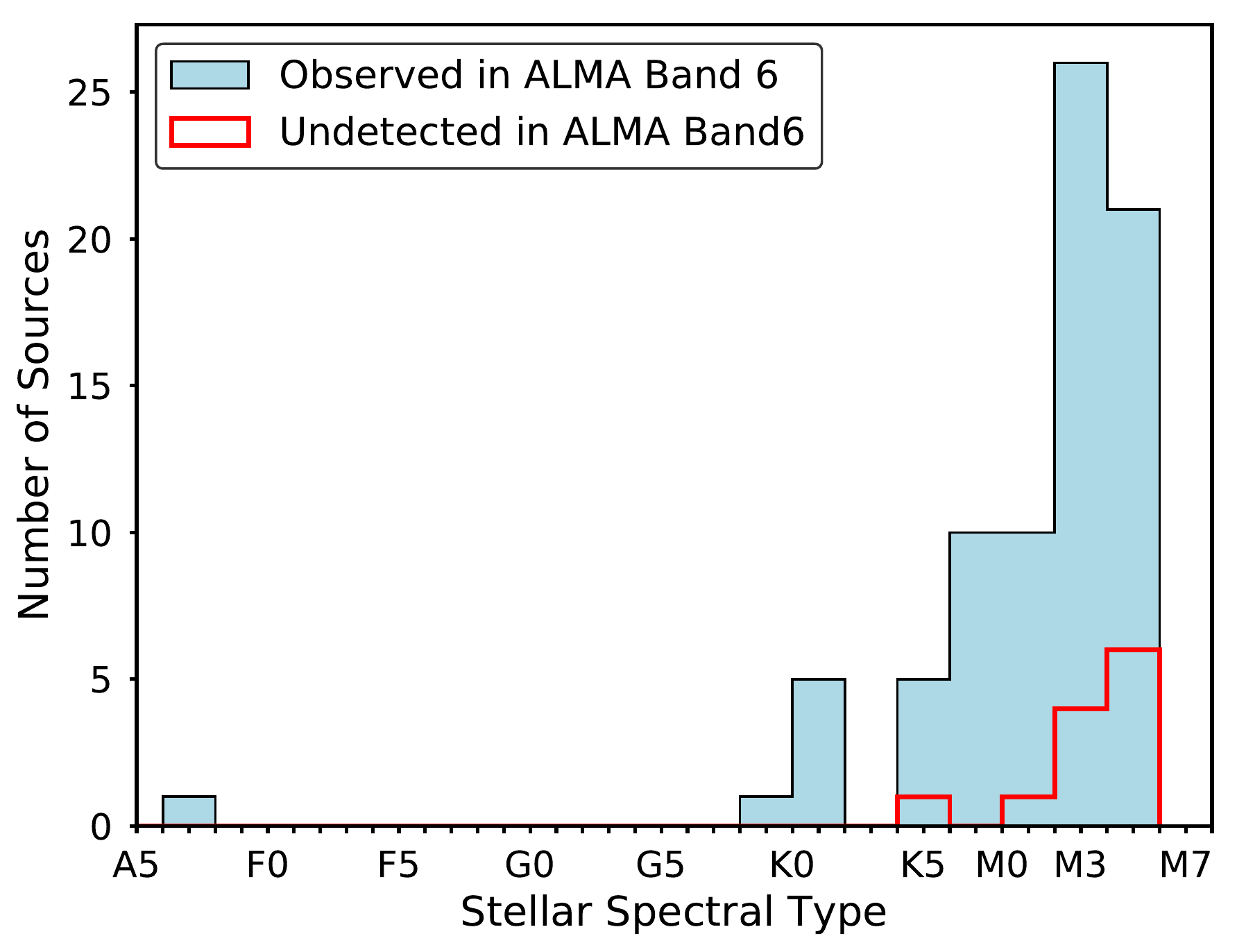}
\caption{\small Distribution of sources in our Lupus sample with known stellar spectral types (Table~\ref{tab-cont}). The blue histogram shows all sources and the red histogram shows the subset of sources undetected in the 1.33~mm continuum by our ALMA Band~6 observations (Section~\ref{sec-cont}).}
\label{Fig1}
\end{centering}
\end{figure}
\capstartfalse

The Lupus complex contains four main star-forming clouds (Lupus I--IV) and is one of the youngest and closest star-forming regions \cite[see review in][]{2008hsf2.book..295C}. Lupus I, III, and IV were observed for the c2d {\it Spitzer} legacy project \citep{2009ApJS..181..321E}, which revealed high disk fractions \cite[70--80\%;][]{2008ApJS..177..551M} consistent with other young disk populations, while Lupus II contains one of the most active nearby T Tauri stars, RU Lup. Lupus is typically assumed to be $\sim$1--3 Myr old \cite[][and references therein]{2008hsf2.book..295C}, but the average age may be as high as $3\pm2$ Myr \citep{2014A&A...561A...2A}. As in Paper~I, we assume that Lupus~III is located at 200~pc, while the other clouds are slightly closer at 150~pc. 

\capstartfalse
\begin{figure*}[!ht]
\begin{centering}
\includegraphics[width=18.4cm]{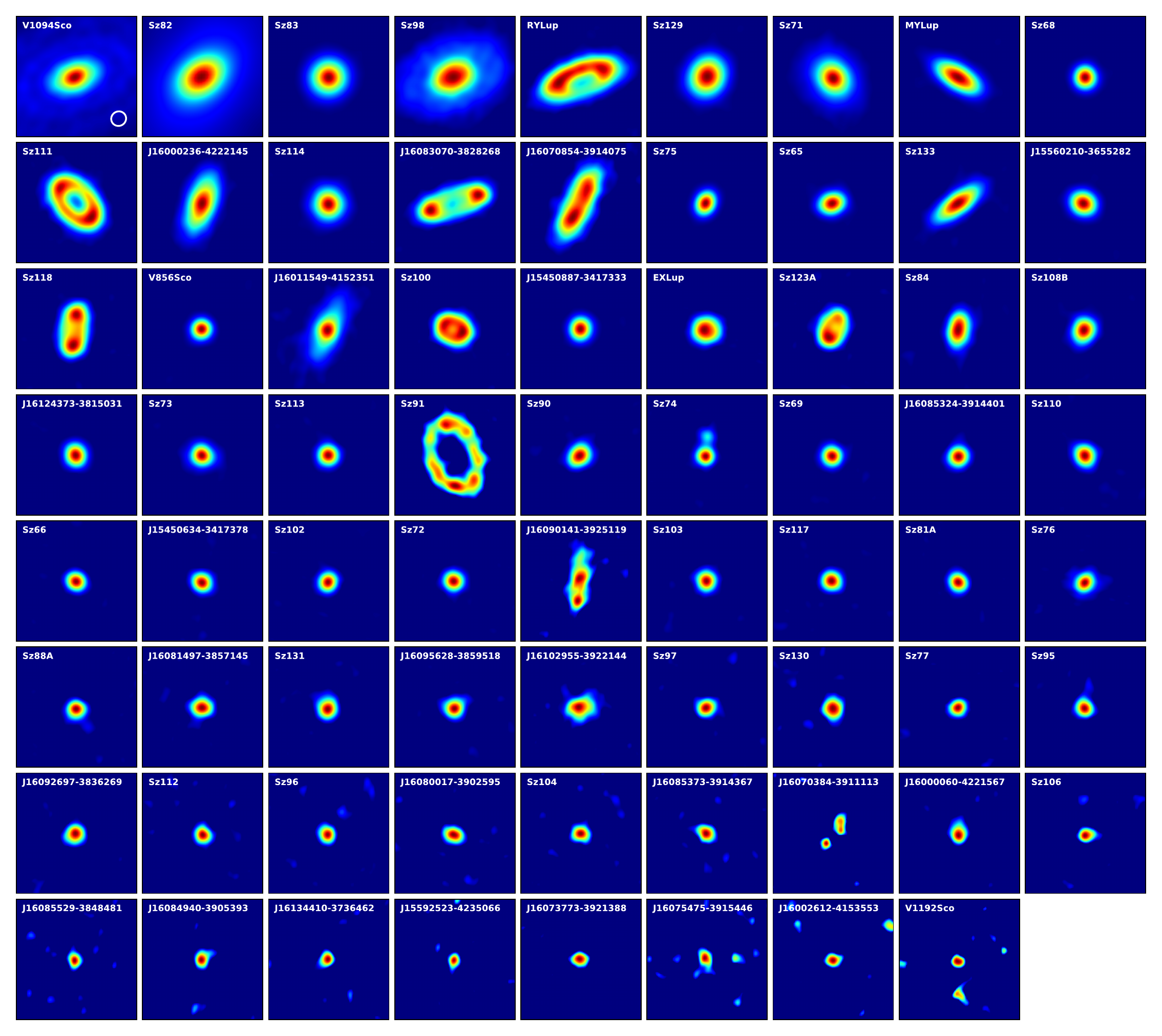}
\caption{\small 1.33~mm continuum images of the 71 Lupus disks detected in our ALMA Band~6 sample, ordered by decreasing continuum flux density (as reported in Table~\ref{tab-cont}). Images are 2\as{}$\times$2\as{} and the typical beam size is shown in the first panel. Each image is scaled so that the maximum is equal to the peak flux and the minimum is clipped at twice the image rms.}
\label{Fig2}
\end{centering}
\end{figure*}
\capstartfalse

Our sample consists of young stellar objects (YSOs) in Lupus  I--IV that are more massive than brown dwarfs (i.e., $M_{\star}>0.1~M_{\odot}$) and host protoplanetary disks (i.e., have Class II or flat IR excess). The preliminary stellar masses used for our sample selection were estimated by fitting absolute $J$-band magnitudes to a 3~Myr \cite{2000A&A...358..593S} model isochrone. Disk classifications were taken from the literature and primarily derived from the IR spectral index slope ($\alpha_{\rm IR}$) between the 2MASS $K_{\rm S}$ (2.2~$\mu$m) and {\it Spitzer} MIPS-1 (24~$\mu$m) bands; for sources without {\it Spitzer} data, disk classifications were approximated from IR excesses and/or accretion signatures (e.g., H$\alpha$~6563\AA{} emission). We do not exclude known binaries, as the binary fraction in Lupus is poorly constrained. 

We identify 95 Lupus members fitting these criteria in the published catalogs of Lupus disks \citep{1994AJ....108.1071H,2011MNRAS.418.1194M,2008ApJS..177..551M,2008hsf2.book..295C,2015ApJS..220...11D,2015A&A...578A..23B}. We note that our sample from Paper~I consisted of 93 sources: in this work, we include four additional sources in our sample (Sz~102, J15560210-3655282, EX~Lup, and Sz~75/GQ~Lup) due to confirmation of Lupus membership via radial velocity (RV) measurements and/or re-classification of disk types based on spectra \citep{2017A&A...602A..33F}. We also remove two sources (J16104536-3854547, J16121120-3832197) as VLT/X-Shooter spectra \citep{2017A&A...600A..20A} have revealed them as background giants due to discrepant surface gravities and RVs \citep{2017A&A...602A..33F}; these two sources were observed, but undetected, in both our ALMA Band~7 and Band~6 observations.

Table~\ref{tab-cont} gives the 95 Lupus disks in our sample and Figure~\ref{Fig1} shows their spectral type distribution. For 76 sources, we provide stellar masses ($M_{\star}$) from \cite{2014A&A...561A...2A} and \cite{2017A&A...600A..20A}, who derived these values using \cite{2000A&A...358..593S} evolutionary models with stellar effective temperatures ($T_{\rm eff}$) and luminosities ($L_{\star}$) estimated from VLT/X-Shooter spectra. We do not provide $M_{\star}$ values for the remaining 19 sources, many of which are obscured with flat IR excesses, complicating the derivation of accurate stellar properties. 

\capstartfalse 
\begin{deluxetable*}{lccccccrcr} 
\tabletypesize{\footnotesize} 
\centering 
\tablewidth{500pt} 
\tablecaption{$1330~\mu$m Continuum Properties \label{tab-cont}} 
\tablecolumns{10}  
\tablehead{ 
 \colhead{Source} 
&\colhead{RA$_{\rm J2000}$} 
&\colhead{Dec$_{\rm J2000}$} 
&\colhead{Dist} 
&\colhead{SpT} 
&\colhead{$M_{\star}$/$M_{\odot}$} 
&\colhead{Ref} 
&\colhead{$F_{\rm cont}$} 
&\colhead{rms} 
&\colhead{$M_{\rm dust}$} \\ 
 \colhead{} 
&\colhead{} 
&\colhead{} 
&\colhead{(pc)} 
&\colhead{} 
&\colhead{} 
&\colhead{} 
&\colhead{(mJy)} 
&\colhead{(mJy beam$^{-1}$)} 
&\colhead{($M_{\oplus}$)} 
} 
\startdata 
Sz~65 & 15:39:27.753 & -34:46:17.577 & 150 & K7.0 & 0.76 $\pm$ 0.18 & 2 & 29.94 $\pm$ 0.20 & 0.12 & 20.24 $\pm$ 0.14 \\
Sz~66 & 15:39:28.264 & -34:46:18.450 & 150 & M3.0 & 0.31 $\pm$ 0.04 & 1 & 6.42 $\pm$ 0.18 & 0.13 & 4.34 $\pm$ 0.12 \\
J15430131-3409153 & 15:43:01.290 & -34:09:15.400 & 150 & ... & ... & ... & -0.24 $\pm$ 0.09 & 0.12 & -0.16 $\pm$ 0.06 \\
J15430227-3444059 & 15:43:02.290 & -34:44:06.200 & 150 & ... & ... & ... & 0.00 $\pm$ 0.09 & 0.10 & 0.00 $\pm$ 0.06 \\
J15445789-3423392 & 15:44:57.900 & -34:23:39.500 & 150 & M5.0 & 0.12 $\pm$ 0.03 & 1 & 0.12 $\pm$ 0.09 & 0.10 & 0.08 $\pm$ 0.06 \\
J15450634-3417378 & 15:45:06.322 & -34:17:38.332 & 150 & ... & ... & ... & 6.18 $\pm$ 0.15 & 0.11 & 4.18 $\pm$ 0.10 \\
J15450887-3417333 & 15:45:08.852 & -34:17:33.835 & 150 & M5.5 & 0.14 $\pm$ 0.03 & 2 & 20.70 $\pm$ 0.18 & 0.12 & 14.00 $\pm$ 0.12 \\
Sz~68 & 15:45:12.849 & -34:17:31.071 & 150 & K2.0 & 2.13 $\pm$ 0.34 & 2 & 66.38 $\pm$ 0.20 & 0.18 & 44.88 $\pm$ 0.14 \\
Sz~69 & 15:45:17.391 & -34:18:28.685 & 150 & M4.5 & 0.19 $\pm$ 0.03 & 1 & 8.05 $\pm$ 0.15 & 0.11 & 5.44 $\pm$ 0.10 \\
Sz~71 & 15:46:44.709 & -34:30:36.054 & 150 & M1.5 & 0.42 $\pm$ 0.11 & 1 & 69.15 $\pm$ 0.31 & 0.12 & 46.76 $\pm$ 0.21
\enddata 
\tablenotetext{}{References: (1) \cite{2014A&A...561A...2A}, (2) \citep{2017A&A...600A..20A}, (3) \cite{2013MNRAS.429.1001A}, (4) \cite{2011MNRAS.418.1194M}, (5) \cite{2008ApJS..177..551M}, (6) \cite{2016ApJ...832..110C}, (7) \cite{2015A&A...578A..23B}, (8) \cite{2008hsf2.book..295C}. Full table available online in machine-readable form.} 
\end{deluxetable*} 
\capstartfalse

 %======================= ALMA OBSERVATIONS  =============================

\section{ALMA OBSERVATIONS}
 \label{sec-observations}

Our ALMA Cycle 3 program (ID: 2015.1.00222.S; PI: Williams) observed 86 sources in our sample in Band~6 on 24 July and 8 September 2016 using 58 12-m antennas on baselines of 15--1110~m and 15--2483~m, respectively. The continuum spectral windows were centered on 234.28 and 216.47~GHz with bandwidths of 2.00 and 1.88~GHz, respectively, for a bandwidth-weighted mean continuum frequency of 225.66~GHz (1.33~mm). On-source integration times were 1.2~min per target for a median continuum rms of 0.10~mJy~beam$^{-1}$. 

The spectral setup also included three windows covering the $^{12}$CO, $^{13}$CO, and C$^{18}$O $J=2$--1 transitions. These spectral windows were centered on 230.51, 220.38, and 219.54~GHz, respectively, with bandwidths of 0.12~GHz, channel widths of 0.24~MHz, and velocity resolutions of 0.16~km~s$^{-1}$. Data were pipeline calibrated by NRAO and included flux, bandpass, and gain calibrations. Flux and bandpass calibrations used observations of J1517-2422 and gain calibrations used observations of J1610-3958. We estimate an absolute flux calibration error of 10\% based on variations in the flux calibrators.

Our ALMA Cycle 4 program (ID: 2016.1.01239.S; PI: van Terwisga) observed an additional seven sources in our sample (Sz~75, Sz~76, Sz~77, Sz~102, EX~Lup, V1094~Sco, J15560210-3655282) in Band~6 on 7 July 2017 (see also van Terwisga, submitted, for descriptions of the ALMA Band~7 observations of these seven sources). The Band~6 array configuration used 44 12-m antennas on baselines of 2600--16700~m. The two continuum spectral windows were centered on 233.99 and 216.49~GHz with bandwidths of 2.00~GHz and 1.88~GHz, respectively, for a bandwidth-weighted mean continuum frequency of 225.52~GHz (1.33~mm). On-source integration times were 2.7~min for a median rms of 0.08~mJy~beam$^{-1}$. The spectral setup included three windows covering the $^{12}$CO, $^{13}$CO, and C$^{18}$O $J=2$--1 transitions. These spectral windows were centered on 230.53, 220.39, and 219.55~GHz, respectively, with bandwidths of 0.12~GHz, channel widths of 0.24~MHz, and velocity resolutions of 0.3~km~s$^{-1}$. NRAO pipeline calibration included flux, bandpass, and gain calibrations using observations of J1427-4206, J1517-2422, and J1610-3958, respectively. We estimate an absolute flux calibration error of 10\% based on variations in the flux calibrators.

The two remaining sources in our sample (Sz~82, Sz~91) have Band~6 continuum observations as well as $^{12}$CO, $^{13}$CO, and C$^{18}$O $J=2$--1 line observations in the ALMA archive (ID: 2013.1.00226.S, 2013.1.01020.S). We downloaded the archival observations and ran the pipeline calibration scripts provided with the data. The observations of Sz~82 (IM~Lup) is published in \cite{2016ApJ...832..110C}. 

Thus we obtain ALMA Band~6 data for all sources in our complete sample of Lupus protoplanetary disks.

 %======================= ALMA RESULTS  ==================================

\section{ALMA RESULTS}
\label{sec-results}

\subsection{1.33 mm Continuum Emission \label{sec-cont}}

We extract continuum images from the calibrated visibilities by averaging over the continuum channels and cleaning with a Briggs robust weighting parameter of $+0.5$; we use $-1.0$ when sources exhibit resolved structure (e.g., for transition disks). The average continuum beam size is 0.25$\times$0.24 arcsec (36$\times$38~au at 150 pc) for our Cycle 3 observations and 0.25$\times$0.21 arcsec (36$\times$32~au at 150 pc) for our Cycle 4 observations. 

We primarily measure continuum flux densities by fitting elliptical Gaussians to the visibility data with {\it uvmodelfit} in CASA. The elliptical Gaussian model has six free parameters: integrated flux density ($F_{\rm cont}$), FWHM along the major axis ($a$), aspect ratio of the axes ($r$), position angle (PA), right ascension offset from the phase center ($\Delta\alpha$), and declination offset from the phase center ($\Delta\delta$). We scale the uncertainties on the fitted parameters by the square root of the reduced $\chi^{2}$ value of the fit. If the ratio of $a$ to its scaled uncertainty is less than five, a point-source model with three free parameters ($F_{\rm cont}$, $\Delta\alpha$, $\Delta\delta$) is fit to the visibility data instead. 

For disks with resolved structure, such as transition disks, flux densities are measured from continuum images using circular aperture photometry. The aperture radius for each source is determined by a curve-of-growth method, in which successively larger apertures are applied until the measured flux density levels off. Uncertainties are then estimated by taking the standard deviation of the flux densities measured within a same-sized aperture placed randomly within the field of view, but away from the source. 

Table~\ref{tab-cont} gives our measured 1.33~mm continuum flux densities and associated uncertainties. The uncertainties are statistical errors and do not include the 10\% absolute flux calibration error (Section~\ref{sec-observations}). Of the 95 sources, 71 are detected with $>3\sigma$ significance; the continuum images of these sources are shown in Figure~\ref{Fig2}. Table~\ref{tab-cont} provides the fitted source centers output by {\it uvmodelfit} for the detections, and the phase centers of our ALMA observations (based on 2MASS source positions) for the non-detections. The image rms for each source, derived from a 4--9\as{} radius annulus centered on the fitted or expected source position for detections or non-detections, respectively, is also given in Table~\ref{tab-cont}.

\subsection{Line Emission \label{sec-line}}

\capstartfalse 
\begin{deluxetable*}{lrrrrr} 
\tabletypesize{\footnotesize} 
\centering 
\tablewidth{500pt} 
\tablecaption{Gas Properties \label{tab-gas}} 
\tablecolumns{6}  
\tablehead{ 
 \colhead{Source} 
&\colhead{$F_{\rm 13CO}$} 
&\colhead{$F_{\rm C18O}$} 
&\colhead{$M_{\rm gas,mean}$} 
&\colhead{$M_{\rm gas,min}$} 
&\colhead{$M_{\rm gas,max}$} \\ 
 \colhead{} 
&\colhead{(mJy~km~s$^{-1}$)} 
&\colhead{(mJy~km~s$^{-1}$)} 
&\colhead{($M_{\rm Jup}$)} 
&\colhead{($M_{\rm Jup}$)} 
&\colhead{($M_{\rm Jup}$)} 
} 
\startdata 
Sz65 & $<102$ & $<60$ & $<1.0$ & ... & ... \\
Sz66 & $<87$ & $<60$ & $<1.0$ & ... & ... \\
J15430131-3409153 & $<84$ & $<51$ & $<1.0$ & ... & ... \\
J15430227-3444059 & $<72$ & $<60$ & $<1.0$ & ... & ... \\
J15445789-3423392 & $<78$ & $<54$ & $<1.0$ & ... & ... \\
J15450634-3417378 & $<84$ & $<57$ & $<1.0$ & ... & ... \\
J15450887-3417333 & 395 $\pm$ 109 & $<54$ & 0.4 & ... & 3.1 \\
Sz68 & $<120$ & $<69$ & $<1.0$ & ... & ... \\
Sz69 & $<81$ & $<45$ & $<1.0$ & ... & ... \\
Sz71 & $<78$ & $<60$ & $<1.0$ & ... & ...
\enddata 
\tablenotetext{}{Full table available online in machine-readable form. Section~\ref{sec-gasmass} explains the derivations of $M_{\rm gas,mean}$, $M_{\rm gas,min}$, and $M_{\rm gas,max}$.} 
\end{deluxetable*} 
\capstartfalse

We extract $^{12}$CO, $^{13}$CO, and C$^{18}$O $J=2$--1 channel maps from the calibrated visibilities by subtracting the continuum and cleaning with a Briggs robust weighting parameter of $+0.5$. Zero-moment maps are created by integrating over the velocity channels showing line emission above the noise. The appropriate velocity range is determined for each source by visual inspection of the channel map and spectrum. If no emission is visible, we sum across the average RV and its dispersion ($\overline{\rm RV} = 2.8 \pm 4.2$~km~s$^{-1}$), as derived for Lupus protoplanetary disks in \cite{2017A&A...602A..33F}.

We measure $^{12}$CO, $^{13}$CO, and C$^{18}$O $J=2$--1 integrated flux densities and associated uncertainties ($F_{\rm 12CO}$, $F_{\rm 13CO}$, and $F_{\rm C18O}$, respectively) from our ALMA zero-moment maps, using the same aperture photometry method described above for structured continuum sources (Section~\ref{sec-cont}). For non-detections, we take upper limits of $3\times$ the uncertainty when using an aperture of the same size as the typical beam (0.25\as{}). 

Of the 95 targets, 48 are detected in $^{12}$CO, 20 in $^{13}$CO, and 8 in C$^{18}$O with $>3\sigma$ significance. All sources detected in C$^{18}$O are also detected in $^{13}$CO, all sources detected in $^{13}$CO are also detected in $^{12}$CO, and all sources detected in $^{12}$CO are also detected in the 1.33~mm continuum. Table~\ref{tab-gas} gives our measured integrated flux densities or upper limits for  $^{13}$CO and C$^{18}$O. We do not provide integrated fluxes for $^{12}$CO because cloud absorption is commonly seen in the spectra (Appendix~\ref{app-co}). Moreover, for nine sources located nearby on the sky in Lupus~III, cloud emission is also seen in the channel maps.

%============================ ANALYSIS ===================================
  
\section{PROPERTIES OF LUPUS DISKS}
\label{sec-analysis}

 %======================= DISK RADII  ==================================

\subsection{Disk Radii \label{sec-radius}}

\capstartfalse 
\begin{deluxetable}{lrrrr} 
\tabletypesize{\footnotesize} 
\centering 
\tablewidth{0pt} 
\tablecaption{Disk Radii \label{tab-radii}} 
\tablecolumns{5}  
\tablehead{ 
 \colhead{Source} 
&\colhead{PA} 
&\colhead{$i$} 
&\colhead{$R_{\rm dust}$} 
&\colhead{$R_{\rm gas}$} \\ 
 \colhead{} 
&\colhead{(deg)} 
&\colhead{(deg)} 
&\colhead{(au)} 
&\colhead{(au)} 
} 
\startdata 
Sz 65 & 108.6 & 61.5 & 64 $\pm$ 2 & 172 $\pm$ 24 \\
Sz 68 & 175.8 & -32.9 & 38 $\pm$ 2 & 68 $\pm$ 6 \\
Sz 71 & 37.5 & -40.8 & 94 $\pm$ 2 & 218 $\pm$ 54 \\
Sz 73 & 94.7 & 49.8 & 56 $\pm$ 3 & 103 $\pm$ 9 \\
Sz 75 & 169.0 & 60.2 & 56 $\pm$ 2 & 194 $\pm$ 21 \\
Sz 76 & 65.0 & -60.0 & 116 $\pm$ 9 & 164 $\pm$ 6 \\
J15560210-3655282 & 55.6 & 53.5 & 56 $\pm$ 2 & 110 $\pm$ 3 \\
Sz 82 & 144.0 & -48.0 & 226 $\pm$ 4 & 388 $\pm$ 84 \\
Sz 84 & 168.0 & 65.0 & 80 $\pm$ 3 & 146 $\pm$ 18 \\
Sz 129 & 154.9 & -31.7 & 68 $\pm$ 2 & 140 $\pm$ 12 \\
RY Lup & 109.0 & 68.0 & 134 $\pm$ 3 & 250 $\pm$ 63 \\
J16000236-4222145 & 160.5 & 65.7 & 112 $\pm$ 3 & 266 $\pm$ 45 \\
MY Lup & 60.0 & 73.0 & 110 $\pm$ 3 & 194 $\pm$ 39 \\
EX Lup & 70.0 & -30.5 & 62 $\pm$ 2 & 178 $\pm$ 12 \\
Sz 133 & 126.3 & 78.5 & 142 $\pm$ 6 & 238 $\pm$ 66 \\
Sz 91 & 17.4 & 51.7 & 154 $\pm$ 4 & 450 $\pm$ 80 \\
Sz 98 & 111.6 & -47.1 & 190 $\pm$ 4 & 358 $\pm$ 52 \\
Sz 100 & 60.2 & 45.1 & 82 $\pm$ 2 & 178 $\pm$ 12 \\
J16083070-3828268 & 107.0 & -74.0 & 182 $\pm$ 4 & 394 $\pm$ 100 \\
V1094 Sco & 110.0 & -55.4 & 334 $\pm$ 20 & 438 $\pm$ 112 \\
Sz 111 & 40.0 & -53.0 & 134 $\pm$ 2 & 462 $\pm$ 96 \\
Sz 123A & 145.0 & -43.0 & 74 $\pm$ 2 & 146 $\pm$ 12
\enddata 
\end{deluxetable} 
\capstartfalse

Disk size is a fundamental property that has been difficult to measure for large samples due to observational constraints. Early measurements of disk radii from sub-mm/mm observations focused on the biggest and brightest disks, perhaps resulting in a common misconception that protoplanetary disks are typically hundreds of au in radius. The recent ALMA surveys of unbiased disk populations have revealed that typical disks are actually closer to a few tens of au in radius, at least in the sub-mm/mm dust \citep{2017A&A...606A..88T, 2017arXiv171104045B}. 

Measuring {\em gas} disk sizes is particularly important because the gas dominates the dynamics of the disk. Gas disk radii are much more difficult to measure, however, due to the faintness of the line emission, especially in the outer regions of the disk. Moreover, the sub-mm/mm dust radius is {\em not} a reliable proxy for the gas disk size because dust grain growth has a radial dependnce and growing dust grains decouple from the gas and drift inward, resulting in the smaller dust disks seen at sub-mm/mm wavelengths \cite[e.g.,][]{2012ApJ...744..162A, 2013A&A...557A.133D, 2016A&A...586A..99H}. 

\capstartfalse
\begin{figure*}
\begin{centering}
\includegraphics[width=18.2cm]{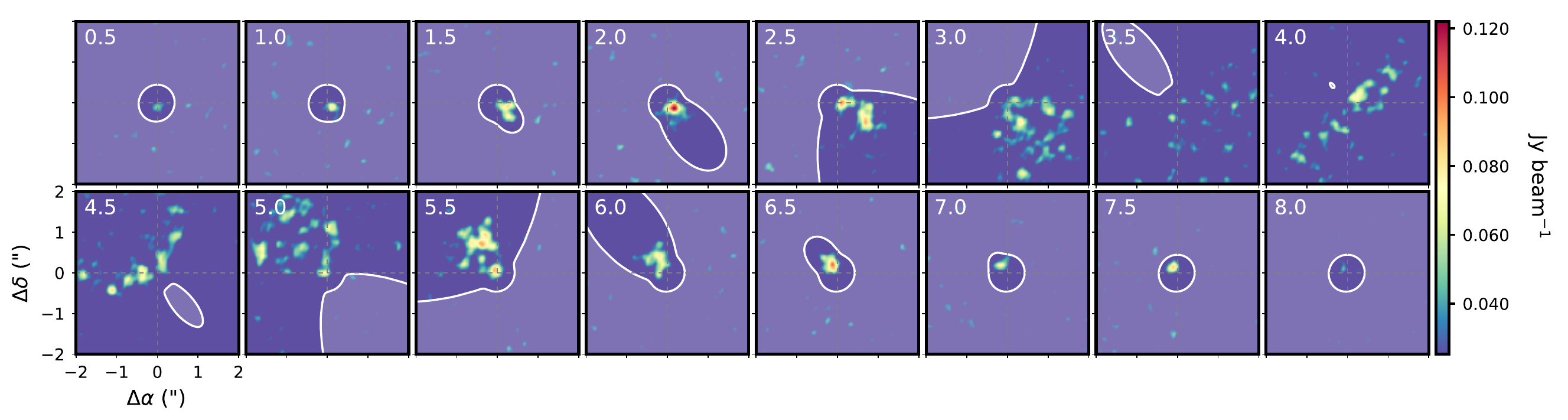}
\caption{\small Keplerian masking applied to the channel maps of Sz~111 (Section~\ref{sec-radius}). In each channel map, only the image regions with expected gas emission from a disk in Keplerian rotation are considered (i.e., the shaded regions are masked out when making the zero-moment map shown in Figure~\ref{Fig4}). The velocities in km~s$^{-1}$ are given in the top left corner of each channel.}
\label{Fig3}
\end{centering}
\end{figure*}
\capstartfalse

\capstartfalse
\begin{figure*}
\begin{centering}
\includegraphics[width=15cm]{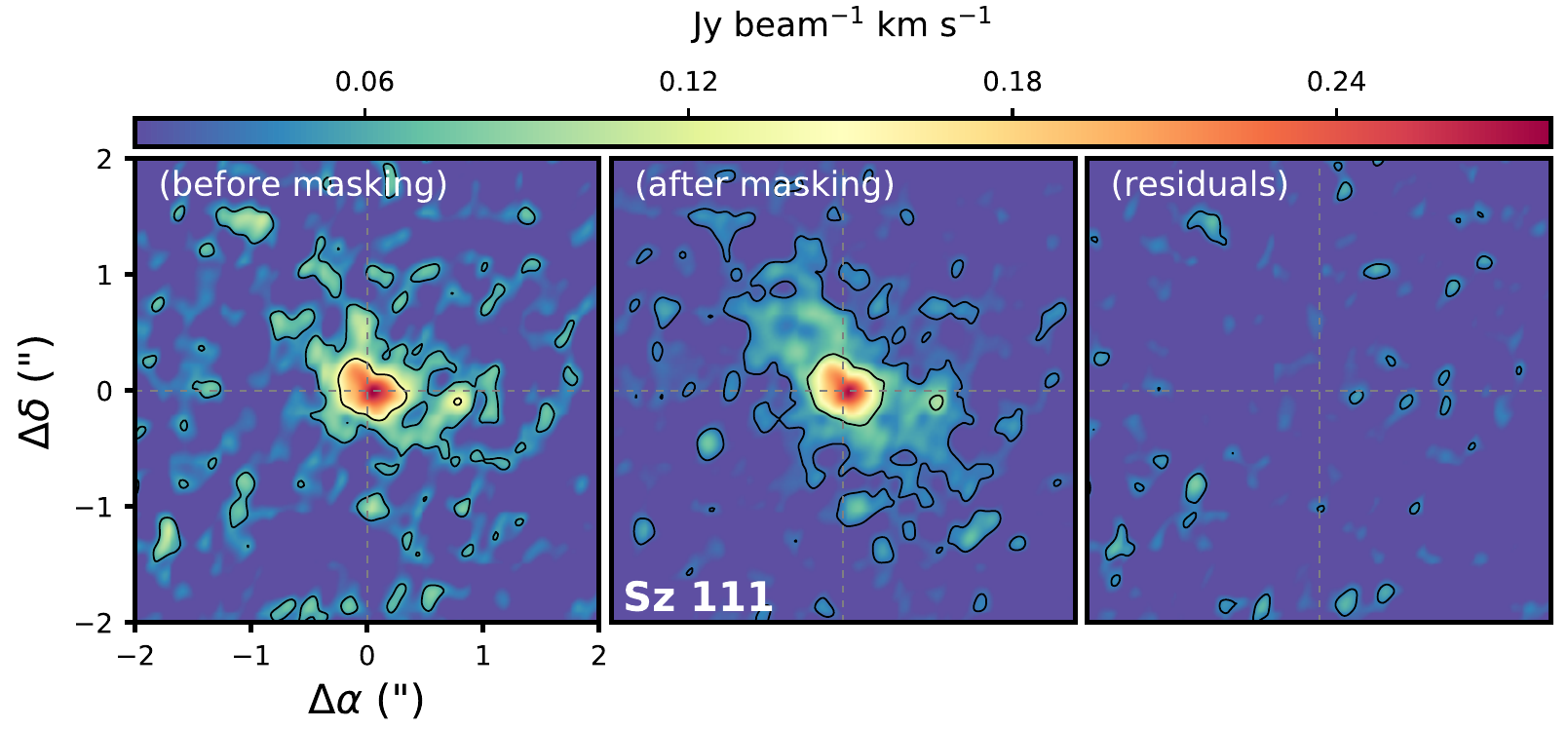}
\caption{\small The zero-moment map of Sz~111 before (left) and after (middle) Keplerian masking as well as the residuals (right). The black lines are 2$\sigma$ and 5$\sigma$ contours, illustrating the improvement in SNR in the fainter outer disk regions (see Section~\ref{sec-radius}). Figure~\ref{FigB2} shows these comparison plots for all Lupus disks with measured $R_{\rm gas}$.}
\label{Fig4}
\end{centering}
\end{figure*}
\capstartfalse

Here we use our ALMA Band~6 data to measure the dust and gas radii of 22 Lupus protoplanetary disks. This is the first large sample of dust and gas radii for disks within a single star-forming region. These disks are listed in Table~\ref{tab-radii} and are selected because they have clearly resolved continuum emission \citep{2017A&A...606A..88T} and exhibit unambiguous $^{12}$CO line emission in multiple velocity channels without significant cloud contamination.  

The dust radii ($R_{\rm dust}$) are measured from the 1.33~mm continuum images using a curve-of-growth method, in which successively larger photometric apertures are applied until the measured flux is 90\% of the total flux. We use elliptical apertures based on the position angle (PA; measured east of north) and inclination ($i$) of the source; these values are mostly taken from \cite{2017A&A...606A..88T}, who derived these parameters using two-layer disk models of the Band~7 continuum visibilities for the full Lupus disks in our sample. For the resolved transition disks with large inner dust cavities (Section~\ref{sec-TD}), we use the PA and $i$ values from \cite{2018arXiv180106154V}, who derived these parameters from by-eye comparisons of the Band~7 first-moment $^{13}$CO maps to model Keplerian velocity fields. The resulting $R_{\rm dust}$ values are given in Table~\ref{tab-radii} in units of au. The errors on $R_{\rm dust}$ are calculated by taking the range of radii within the uncertainties on the 90\% flux measurement. Comparing our $R_{\rm dust}$ values to the $R_{\rm out}$ values derived in \cite{2017A&A...606A..88T} for the 12 sources common to both samples shows good agreement despite the very different analysis methods: the average ratio is 1.06 with a standard deviation of 0.37.

The gas radii ($R_{\rm gas}$) are measured from the $^{12}$CO zero-moment maps using the same curve-of-growth method described above for the $R_{\rm dust}$ measurements (we note that this method is robust against the effects of cloud absorption, as this only affects the blue- {\em or} red-shifted side of the disk emission). The same PA and $i$ values used for measuring $R_{\rm dust}$ are used again for measuring $R_{\rm gas}$, which is important because applying different $i$ values can lead to significantly different radii when implementing elliptical apertures in a curve-of-growth analysis. 

Before measuring $R_{\rm gas}$, we also re-construct the zero-moment maps using a ``Keplerian masking" technique to increase the signal-to-noise ratio (SNR), especially in the fainter outer disk regions (see the Appendixes in \citealt{2017A&A...606A.125S} and Trapman et al., in prep, for detailed descriptions of the Keplerian masking technique). In short, Keplerian masking takes advantage of the fact that the gas disk is in Keplerian rotation, and thus will only emit in certain regions of the sky in a given velocity channel. Masking the pixels outside of these regions in each velocity channel therefore enhances the SNR in the final integrated zero-moment map. We show the Keplerian masking technique applied to Sz~111 in Figure~\ref{Fig3}, and also the improvement in the SNR in the outer disk regions in Figure~\ref{Fig4} (plots for all 22 disks are shown in Figure~\ref{FigB3}). The resulting $R_{\rm gas}$ values are given in Table~\ref{tab-radii}.

We find that the gas disks are universally larger than the dust disks, by an average factor of $1.96\pm0.04$ (where this is the mean and standard error on the mean). We note that this result holds even when using a 68\% (rather than 90\%) flux threshold for the radius measurements as well as when using circular (rather than elliptical) apertures (see Appendix~\ref{app-radii}.). Although previous observations of large individual disks have shown gas disks extending beyond dust disks by similar factors \cite[e.g.,][]{2007A&A...469..213I, 2009A&A...501..269P, 2012ApJ...744..162A, 2013A&A...557A.133D, 2016ApJ...832..110C}, this is the first indication of systematically larger gas radii in a coherent population of disks. We note that our results differ from those of \cite{2017arXiv171104045B}, who found in a sample of seven Upper Sco disks only four with larger gas radii, and no clear overall trend (see their Figure 6); however, their sample is much smaller and they did not apply Keplerian masking. We discuss the implications in Section~\ref{sec-radii}.

 %======================= DISK MASSES  ==================================

\subsection{Disk Masses \label{sec-mass}}

\subsubsection{Dust Masses \label{sec-dustmass}}
 
Dust emission at sub-mm/mm wavelengths is typically optically thin in most regions of a protoplanetary disk, in which case estimates of dust mass can be directly obtained from measurements of the sub-mm/mm continuum flux. \cite{2017arXiv171008426R} calculated the spectral index from far-IR to mm wavelengths for 284 protoplanetary disks, showing that the spectral index distributions become remarkably similar from 880~$\mu$m to 5~mm, despite the significant range in wavelength, which indeed suggests that the overall disk dust emission is generally optically thin at these longer wavelengths.

\begin{figure}
\begin{centering}
\includegraphics[width=8.5cm]{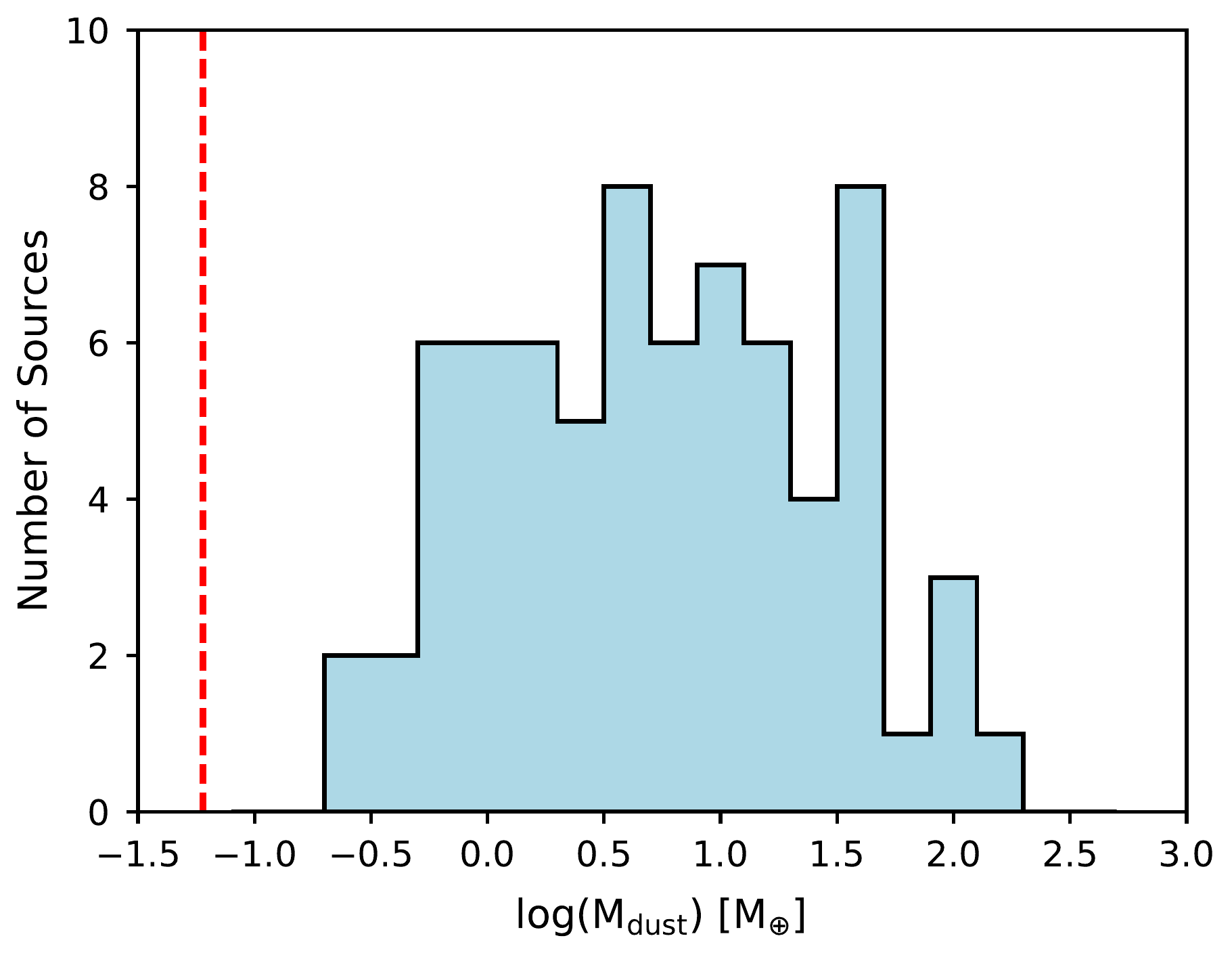}
\caption{\small Distribution of dust masses ($M_{\rm dust}$) for the Lupus disks detected in the 1.33~mm continuum (Section~\ref{sec-dustmass}). The dashed red line shows the $3\sigma$ upper limit from the stacked continuum non-detections (Section~\ref{sec-stacks}); the stark contrast to the faintest continuum detection suggests protoplanetary disks evolve rapidly to debris disk levels once clearing begins.}
\label{Fig5}
\end{centering}
\end{figure}

Assuming dust emission at sub-mm/mm wavelengths is also isothermal, the sub-mm/mm continuum flux from a protoplanetary disk at a given wavelength ($F_{\nu}$) can be directly related to the mass of the emitting dust ($M_{\rm dust}$), as shown in \cite{1983QJRAS..24..267H}:

\begin{equation}
M_{\rm dust}=\frac{F_{\nu}d^{2}}{\kappa_{\nu}B_{\nu}(T_{\rm dust})} \approx 2.03\times10^{-6} \left(\frac{d}{150}\right)^{2}F_{1.33 {\rm mm}},
\label{eqn-mass}
\end{equation}

where $B_{\nu}(T_{\rm dust})$ is the Planck function for a characteristic dust temperature of $T_{\rm dust}=20$~K, the median for Taurus disks \citep{2005ApJ...631.1134A}. We take the dust grain opacity, $\kappa_{\nu}$, as 10 cm$^{2}$ g$^{-1}$ at 1000 GHz and use an opacity power-law index of $\beta_{\rm d} =1.0$ \citep{1990AJ.....99..924B}. Distances, 1.33~mm continuum flux densities, and associate uncertainties are from Table~\ref{tab-cont}. The calculated $M_{\rm dust}$ values are given in Table~\ref{tab-cont} and the $M_{\rm dust}$ distribution is shown in Figure~\ref{Fig5}. The median fractional difference between the dust masses derived here from our Band~6 data compared to the values derived in Paper~I from our Band~7 data is 15\%. 

As in previous works \cite[e.g., ][]{2013ApJ...771..129A, 2016ApJ...828...46A, 2016ApJ...827..142B, 2016ApJ...831..125P, 2017AJ....153..240A}, we also fit the $M_{\rm dust}$--$M_{\star}$ relation using the Bayesian linear regression method of \cite{2007ApJ...665.1489K} to take into account upper limits, error bars on both axes, and intrinsic scatter in the data. Using the same Monte Carlo method from Paper~I to account for the 19 sources with unknown stellar masses (Section~\ref{sec-sample}), we find the relation:

\begin{equation} 
{\rm log} (M_{\rm dust} ) = 1.3 (\pm0.2) + 1.8 (\pm0.3)  \times {\rm log} (M_{\star}),
\label{eqn-mdms}
\end{equation}

with a dispersion of $0.8 \pm 0.2$ dex. These fitted parameters are nearly identical to (and well within the errors of) those derived from our Band~7 observations in Paper~I. We note that our assumption of an isothermal disk temperature could effect the fitted $M_{\rm dust}$--$M_{\star}$ relation, if there is a dependence of $T_{\rm dust}$ on stellar parameters. Although \cite{2013ApJ...771..129A} derived the relation $T_{\rm dust}=25 {\rm K} \times (L_{\star}/L_{\odot})^{0.25}$ using two-dimensional continuum radiative transfer models, recent ALMA observations suggest that $T_{\rm dust}$ is actually largely independent of stellar parameters. In particular, \cite{2017A&A...606A..88T} used detailed modeling of 36 resolved Lupus disks to show a lack of correlation between $T_{\rm dust}$ and $L_{\star}$ or $M_{\star}$, at least for their sample. Thus applying model-derived relations runs the risk of introducing artificial correlations or increasing the dispersion of true correlations. Indeed, applying the $T_{\rm dust}$--$L_{\star}$ relation derived by \cite{2013ApJ...771..129A} to ALMA disk surveys results in a shallower slope when compared to assuming an isothermal disk temperature of $T=20$~K \citep{2016ApJ...831..125P}.

\capstartfalse
\capstartfalse

\capstartfalse 
\begin{deluxetable*}{lccrrrc} 
\tabletypesize{\footnotesize} 
\centering 
\tablewidth{500pt} 
\tablecaption{Secondary Source Properties \label{tab-secondary}} 
\tablecolumns{7}  
\tablehead{ 
 \colhead{Source} 
&\colhead{RA$_{\rm J2000}$} 
&\colhead{Dec$_{\rm J2000}$} 
&\colhead{$F_{\rm cont }$ (mJy)} 
&\colhead{$F_{\rm cont}$ (mJy)} 
&\colhead{PA (deg)} 
&\colhead{$\rho$ (arcsec)} \\ 
 \colhead{[primary]} 
&\colhead{[secondary]} 
&\colhead{[secondary]} 
&\colhead{[secondary]} 
&\colhead{[primary]} 
&\colhead{} 
&\colhead{} 
} 
\startdata 
Sz~68                           & 15:45:12.646  &  -34:17:29.768     & $3.35\pm0.10$   & $66.38\pm0.20$   &  297.32 &   2.84  \\
Sz~74                           & 15:48:05.212  &  -35:15:53.032     &                            & $11.51\pm0.34$   & 355.96  &    0.31  \\
Sz~81A                         & 15:55:50.317  &  -38:01:32.262     &  $1.45\pm0.10$  &  $4.24\pm0.11$    & 18.61    &   1.93  \\ 
J16070384-3911113     & 16:07:03.585  &  -39:11:12.022     &   $0.38\pm0.10$  &  $0.98\pm0.29$  & 267.54  &    2.84  \\
Sz~88A                         & 16:07:00.567  &  -39.02.20.202     &                            &  $3.72\pm0.11$   &  212.61 &   0.34  \\
J16073773-3921388    & 16:07:37.562  &  -39:21:39.218     &  $0.45\pm0.10$   &  $0.52\pm0.09$   & 266.61  &    1.72  \\
V856~Sco                     & 16:08:34.390  &  -39:06:19.310    &  $8.21\pm0.10$   &  $23.03\pm0.11$  & 112.68 &    1.45
\enddata 
\end{deluxetable*} 
\capstartfalse

 \subsubsection{Gas Masses \label{sec-gasmass}}

We estimate bulk gas masses using our CO isotopologue line observations, following the methods of \cite{2014ApJ...788...59W} and \cite{2016ApJ...828...46A}. In short, \cite{2014ApJ...788...59W} used parameterized gas disk models to show that combining the $^{13}$CO and C$^{18}$O isotopologue lines, with their moderate-to-low optical depths, provides a relatively simple and robust proxy of bulk gas content in protoplanetary disks, except for exceptionally cold or low-mass disks \citep{2016A&A...594A..85M}. In Paper~I, we applied this method to protoplanetary disks in Lupus by comparing our Band~7 observations of the $^{13}$CO and C$^{18}$O $J=3$--2 line luminosities to the WB14 model grids. We considered both the ISM C$^{18}$O isotopologue abundance as well as a factor of 3$\times$ lower in order to take into account CO isotope-selective photodissociation \citep{1988ApJ...334..771V}, which affects CO-derived gas masses \citep{2016A&A...594A..85M, 2017A&A...599A.113M}.

Here we apply the same method to derive bulk gas masses using our Band~6 observations of the $^{13}$CO and C$^{18}$O $J=2$--1 line luminosities. Our derived gas masses are given in Table~\ref{tab-gas}. For the 8 sources detected in both $^{13}$CO and C$^{18}$O, we calculate the mean (in log space) of the WB14 model grid points within $\pm$3$\sigma$ of our measured $^{13}$CO and C$^{18}$O line luminosities ($M_{\rm gas}$), and also set upper ($M_{\rm gas,max}$) and lower ($M_{\rm gas,min}$) limits based on the maximum and minimum WB14 model grid points consistent with the data. For the 12 disks with $^{13}$CO detections and C$^{18}$O upper limits, we similarly calculate $M_{\rm gas}$ and $M_{\rm gas,max}$ but set $M_{\rm gas,min}$ to zero as isotope-selective photodissociation may be stronger for low-mass disks \citep{2016A&A...594A..85M}. For the 75 disks undetected in both lines, we set only upper limits to the gas masses using the maximum model grid points consistent with the $^{13}$CO and C$^{18}$O line luminosity upper limits. We note that the \cite{2014ApJ...788...59W} model grid only explores radii from 30 to 200~au, thus these gas non-detections may be due to small gas disks; however, assuming purely optically thick emission and a minimum CO temperature of 20~K, we estimate that our observations should have been able to detect all disks greater than $\sim$30~au in diameter, comparable to our beam size.

Within the uncertainties, the $M_{\rm gas}$ values derived in this work from our Band~6 ($J=2$--1) data are consistent with those derived from our Band~7 ($J=3$--2) data in Paper~I for the sources detected in $^{13}$CO and C$^{18}$O in both surveys. However, the uncertainties are large and the sample is small (only five sources, three of which are transition disks with resolved dust cavities). Moreover, the $M_{\rm gas}$ values derived from the Band~6 data are systematically $\sim$0.3--0.5~dex higher than those derived from the Band~7 data. For these sources, $M_{\rm gas} \ge 10^{-3} M_{\odot}$, thus they are unlikely to be affected by isotope-selective photodissociation \cite[see Figure 7 in][]{2016A&A...594A..85M}

\subsection{Stacked Analysis \label{sec-stacks}}

We perform a stacking analysis to constrain the average dust and gas masses of the individually undetected sources. Before stacking the non-detections, we center each image on the expected source location and normalize the flux to 150~pc. The flux densities are then measured in the stacked images using circular aperture photometry, as in Section~\ref{sec-cont}. We confirm that the source locations are known to sufficient accuracy for stacking by measuring the average offset of the detected sources from their phase centers: we find $\langle\Delta\alpha\rangle=-0.11$\as{} and $\langle\Delta\delta\rangle=-0.18$\as{}, both smaller than the beam size. 

We first stack the 24 continuum non-detections, but do not find a significant mean signal in the continuum, $^{12}$CO, $^{13}$CO, or C$^{18}$O stacks. The lack of line emission is expected given the undetected continuum, but the absence of continuum emission is surprising given the sensitivity of the stacked image. Using an aperture the same size as the beam, we measure a mean signal of $0.00\pm0.04$~mJy; we can confirm this by calculating the mean and standard error on the mean from the continuum fluxes reported in Table~\ref{tab-cont}, which similarly gives $-0.06\pm0.03$~mJy. This translates to a 3$\sigma$ upper limit on the average dust mass of individually undetected continuum sources of $\sim$5 Lunar masses (0.06~$M_{\oplus}$), comparable to debris disk levels \citep{2008ARA&A..46..339W}. The stark contrast between the detections and stacked non-detections, illustrated in Figure~\ref{Fig5}, suggests that protoplanetary disks evolve rapidly to debris disk levels once disk clearing begins \citep{2014prpl.conf..475A}. 

We then stack the 12 sources detected in the continuum and $^{13}$CO, but not C$^{18}$O. We measure a continuum mean signal of $42.10\pm0.64$ mJy and a $^{13}$CO mean signal of $1030\pm150$~mJy~km~s$^{-1}$. The stacking also reveals a marginally significant mean signal for C$^{18}$O of $270\pm90$~mJy~km~s$^{-1}$. The stacked continuum flux corresponds to $M_{\rm dust}\sim28~M_{\oplus}$ and the stacked line fluxes correspond to $M_{\rm gas}\sim0.36~M_{\rm Jup}$, giving an average gas-to-dust ratio of only $\sim$4 for sources detected in the continuum and $^{13}$CO, but not C$^{18}$O.

Finally, we stack the 51 sources detected in the continuum, but undetected in $^{13}$CO and C$^{18}$O. We measure a continuum mean signal of $19.06\pm0.12$~mJy. The stacking also reveals marginally significant mean signals for $^{13}$CO and C$^{18}$O; the stacked gas fluxes are $190\pm50$~mJy~km~s$^{-1}$ and $40\pm10$~mJy~km~s$^{-1}$, respectively. Note that these stacked line fluxes of the non-detections are significantly lower than the line fluxes of the detections, similar to what is seem for the continuum. The stacked continuum flux corresponds to $M_{\rm dust}\sim13~M_{\oplus}$, while the stacked line fluxes correspond to $M_{\rm gas}\sim0.14~M_{\rm Jup}$ for an average gas-to-dust ratio of just $\sim$3 for disks detected in the continuum but undetected in $^{13}$CO and C$^{18}$O.

 %======================= INDIVIDUAL SOURCES  ==================================

 \subsection{Individual Sources \label{sec-individual}}

 \subsubsection{Transition Disks \& Asymmetric Disks \label{sec-TD}}

\cite{2018arXiv180106154V} identified 11 transition disks with large ($>20$~au) inner dust cavities in our Band~7 survey from Paper~I. At the spatial resolution of those observations ($\sim$$0\farcs35$), half of the transition disks have cavities clearly resolved in their Band~7 continuum images, while the other half are only marginally resolved and primarily identified through the nulls in their Band~7 continuum visibility curves. The higher-resolution ($\sim$$0\farcs25$) Band~6 data presented in this work now clearly resolve all of these cavities in the continuum image plane. Furthermore, with the higher continuum sensitivity of our Band~6 data, two additional disks (J16090141-3925119, J16070384-3911113) now show evidence for dust cavities with radii of $\sim$30~au in their Band~6 data; neither of these disks have been previously identified in the literature as transition disk candidates by their spectral energy distribution (SED) shapes. In this work, we refer to these 13 disks as ``resolved" transition disks. The ``unresolved" transition disks in our sample are those identified by their SEDs, but without resolved dust cavities in current sub-mm/mm continuum data.

Another interesting aspect of some resolved transition disks is the appearance of azimuthal asymmetries when observed at high spatial resolution: both RY~Lup and Sz~123A appear to be azimuthally asymmetric, with contrasts of 2.2 and 1.2, respectively. Extreme azimuthal asymmetries have been observed in several other resolved transition disks, usually linked to dust trapping in vortices \cite[e.g.,][]{2013Sci...340.1199V,2015ApJ...812..126C, 2017ApJ...848L..11K}, whereas shallower asymmetries with contrasts of $<2$--3 \cite[e.g.,][]{2014ApJ...783L..13P, 2015A&A...584A..16P} have been explained by other mechanisms \cite[e.g., eccentricity;][]{2013A&A...553L...3A}.

 \subsubsection{Secondary Sources \label{sec-secondary}}

We detect seven secondary sources that are not accounted for in our target list. The coordinates and 1.33~mm fluxes ($F_{\rm cont}$) of these secondary sources are given in Table~\ref{tab-secondary}. Additionally, we provide the position angle (PA; measured east of north) and projected angular separation ($\rho$) of each secondary source relative to its primary source. The $F_{\rm cont}$ values are estimated by fitting point source models to the visibility data with {\it uvmodelfit}, as in Section~\ref{sec-cont}, and are consistent with values obtained with aperture photometry. We do not provide $F_{\rm cont}$ values for the secondary sources to Sz~74 and Sz~88A, as they are too close to their primary source to allow reliable individual flux measurements. 

Sz~68, Sz~74, Sz~81A, and V856~Sco are known binary stars from the literature \citep{1993A&A...278...81R, 1997A&A...318..472L, 2001A&A...369..249W} and here we detect the disks of their secondary companions. The PA and $\rho$ values measured from our mm detections match those reported in the literature for the stellar binary systems.

Sz~88A shows a very close ($0\farcs38$) secondary mm component, which is not immediately visible in Figure~\ref{Fig2} due to the relative brightness of the primary target. Although this star has a known companion at 1.5\as \cite[Sz~88B;][]{1993A&A...278...81R}, which we do not detect in our data, follow-up with VLT/NACO did not reveal any closer companions \citep{2006A&A...459..909C} that could be the possible host star for this potential disk.

\begin{figure}
\begin{centering}
\includegraphics[width=8.5cm]{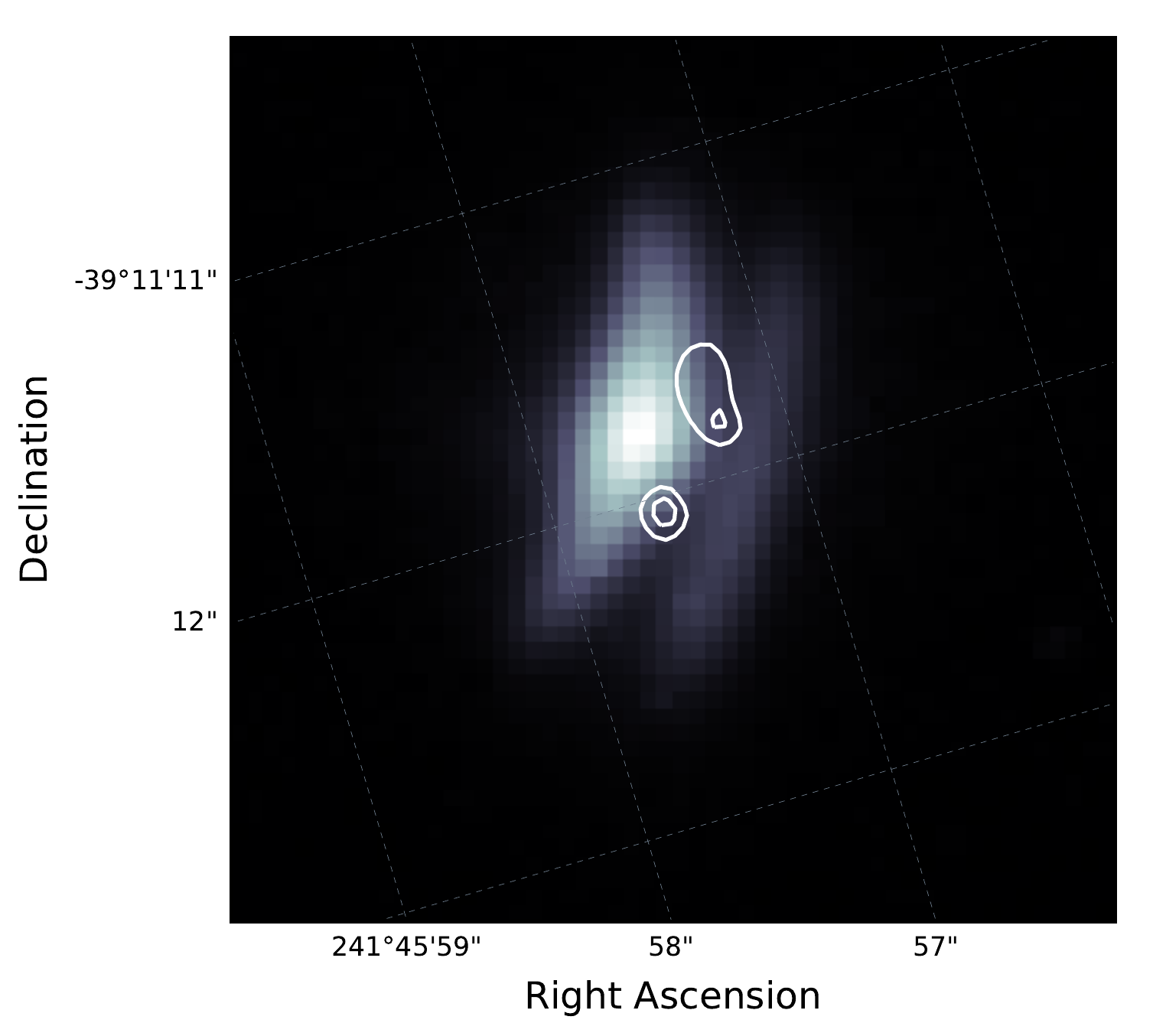}
\caption{\small Hubble/ACS image (PI: Stapelfeldt) of J16070384-3911113, revealing an edge-on flared disk. The white lines are 3$\sigma$ and 4$\sigma$ contours of the ALMA 1.33~mm continuum emission, which align with the disk mid-plane. This suggests that the two mm points are the bright edges of an edge-on disk, which is possibly a transition disk (Section~\ref{sec-TD}).}
\label{Fig6}
\end{centering}
\end{figure}

J16070384-3911113 shows a weak companion at a projected separation of 2.85\as{} but there are no known stellar sources near this location reported in the literature. Although this source also appears to be a close binary in our Band~6 data (see Figure~\ref{Fig2}), the two mm continuum points are likely just the bright limbs of an edge-on disk. This interpretation is supported by its publicly available Hubble/ACS optical image (Proposal ID: 14212; PI: Stapelfeldt), which reveals a flared edge-on disk with a mid-plane that is aligned with the two mm continuum points, as shown in Figure~\ref{Fig6}. Additionally, the detected $^{12}$CO emission in our Band~6 data and $^{13}$CO emission in our Band~7 data both show Keplerian rotation encompassing both continuum points. The edge-on nature of the disk also explains its apparently flat IR excess.

J16073773-3921388 has a visual companion detected in the optical at 3.2\as{} to the north \citep{2008ApJS..177..551M}. However, the sub-mm/mm component we detect is at 1.76\as{} to the west, making it unlikely to be the same object (if bound) given the time elapsed between the observations.

\capstartfalse
\capstartfalse

 \subsubsection{Outflow Sources \label{sec-outflows}}

Two disks in our survey, Sz~83 and J15450634-3417378, exhibit unusual structures in $^{12}$CO emission that may indicate a wide-angle outflow or remnant thereof. Namely, their channel maps show off-center rings toward each source (see Appendix~\ref{app-outflow}), and the coherence in position and velocity shows that these are related to the YSO and are not cloud confusion. Sz~83 is the famous source RU~Lup, one of the most active T Tauri stars in Lupus with known outflows and jets \cite[e.g.,][]{2005AJ....129.2777H}. J15450634-3417378 is lesser known, detected previously at sub-mm/mm continuum wavelengths but lacking any previous evidence of outflows.

The nature of these features is not known. Interestingly, both are slightly offset from the systemic velocity of the disk. One possibility is that they are slow moving flows from the outer regions of the disk. Such disk winds, magnetically launched from several au radii, have been theorized in non-ideal MHD models \citep{2015ApJ...801...84G}, including even one-sided flows \citep{2017ApJ...845...75B}, and have been observed in recent ALMA observations \citep{2016Natur.540..406B, 2017NatAs...1E.146H}. An alternative possibility is that the flows are remnants of eruptive FUOr-like events, in which similarly large-scale, slow-moving arc-like structures are found \citep{2017MNRAS.465..834Z, 2017MNRAS.466.3519R, 2017MNRAS.468.3266R, 2018MNRAS.473..879P}.

One other source, J15430131-3409153, also shows extended $^{12}$CO emission in its channel maps that appears to be aligned with the position of the YSO (see Figure~\ref{FigC3}), which we do {\em not} detect in the 1.33~mm continuum. However, the $^{12}$CO emission is actually associated with the nearby outflow source IRAS~15398-3359 \cite[e.g.,][]{2013ApJ...779L..22J}, which is located at the outer edges of the field of view of J15430131-3409153.

\capstartfalse
\begin{figure*}
\begin{centering}
\includegraphics[width=18.2cm]{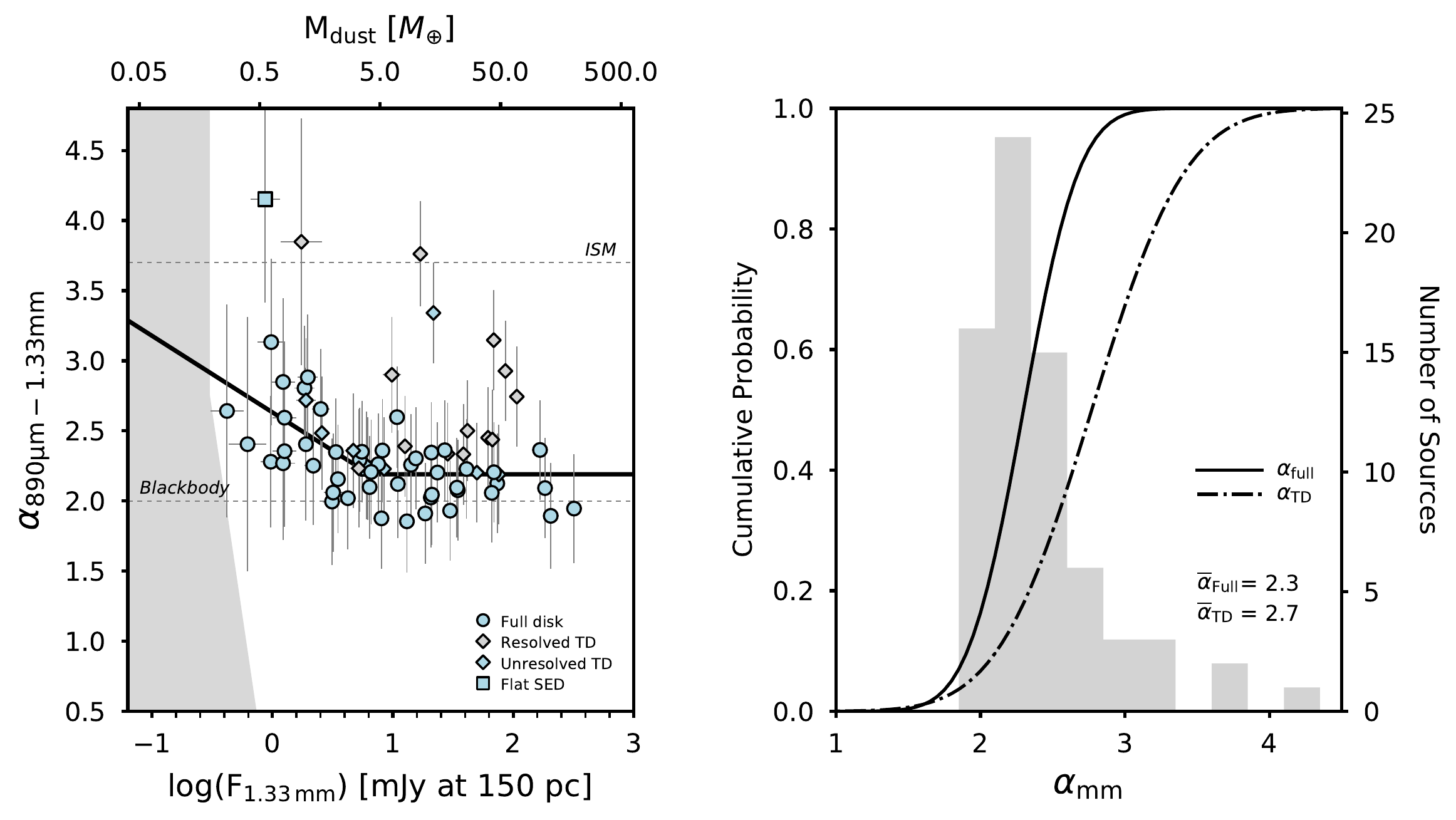}
\caption{\small {\bf Left:} The mm spectral index ($\alpha_{\rm mm}$) versus 1.3~mm flux ($F_{\rm 1.33~mm}$) for Lupus disks, where $\alpha_{\rm mm}$ is calculated between 890~$\mu$m (Band~7) and 1.33~mm (Band~6), and $F_{\rm 1.33~mm}$ is normalized to 150~pc (Section~\ref{sec-alpha}). Blue circles are full disks, blue squares are sources with flat IR excess, and blue diamonds are unresolved transitions disks; gray diamonds are transition disks with resolved cavities (Section~\ref{sec-TD}). The black line gives a piecewise linear fit to the full disks (see Equation~\ref{eqn-alpha}). The shaded region shows the 3$\sigma$ upper limits, illustrating that our results are not due to observational biases. Dashed lines show $\alpha_{\rm mm}$ values for the ISM and pure blackbody emission. The top axis gives approximate dust masses ($M_{\rm dust}$) based on Equation~\ref{eqn-mass}. {\bf Right:} cumulative distributions of $\alpha_{\rm mm}$ values for different subsets of the Lupus disk population, over-plotted on a histogram of all $\alpha_{\rm mm}$ values. The solid and dashed-dotted lines show full disks and transition disks, respectively.}
\label{Fig7}
\end{centering}
\end{figure*}
\capstartfalse

%===============================  DISCUSSION ============================

\section{DISCUSSION}
\label{sec-discussion}

%=========================== GRAIN GROWTH ============================

 \subsection{Dust Grain Growth \label{sec-alpha}}

One of the largest uncertainties in converting sub-mm/mm continuum flux into disk dust mass (e.g., Equation~\ref{eqn-mass}) is the power-law index of the dust opacity spectrum, $\beta_{\rm d}$, where $\kappa_{\nu} \propto \nu ^ {\beta_{\rm d}}$. If the disk dust emission is optically thin and in the Rayleigh-Jeans regime, its sub-mm/mm SED has a power-law dependence on frequency, such that $F_{\nu} \propto \kappa_{\nu} \nu^2 \propto \nu^{2+\beta_{\rm d}}$. In this case, we can fit the observed sub-mm/mm SED between two frequencies with $F_{\nu_1}/F_{\nu_2} = \left( \nu_1 / \nu_2 \right) ^ {\alpha_{\rm mm}}$, where $\alpha_{\rm mm}$ is the sub-mm/mm spectral index, and then derive the dust opacity index using $\beta_{\rm d} = \alpha_{\rm mm} - 2$. For ISM dust, $\beta_{\rm d}\approx1.7$ \citep{2001ApJ...554..778L}, a value that should decrease (i.e., become more grey) as grains grow \cite[e.g.,][]{2006ApJ...636.1114D}.

Here we derive $\alpha_{\rm mm}$ between 890~$\mu$m and 1.33~mm for the 70 Lupus disks detected in both Band~6 (this work) and Band~7 (Paper~I; van Terwisga et al., in prep). We use the standard equation for deriving $\alpha_{\rm mm}$ described above, and calculate uncertainties by propagating the errors on the flux measurements, which include both the statistical error and the 10\% flux calibration uncertainty (Section~\ref{sec-observations}). Our results are given in Figure~\ref{Fig7}, which shows $\alpha_{\rm mm}$ as a function $F_{\rm 1.33~mm}$ (normalized to 150~pc) as well as the cumulative distributions of the $\alpha_{\rm mm}$ values. The median $\alpha_{\rm mm}$ value is 2.25 (when excluding transition disks with resolved cavities; see below), similar to what is seen in other young regions at these wavelengths \citep{2017arXiv171008426R} and also at slightly longer wavelengths of 1--3~mm \citep{2014prpl.conf..339T}. Moreover, these $\alpha_{\rm mm}$ values translate to $\beta_{\rm d}$ values much lower than that of the ISM, implying significant grain growth in Lupus disks.

We note that for blackbody emission, $F_{\nu} \propto \nu^2$, thus $\alpha_{\rm mm}>2$ is the limit for optically thin, grey body emission in the Rayleigh-Jeans regime. However, disks may exhibit $\alpha_{\rm mm} < 2$ when they deviate from these conditions, for example in the case of exceptionally cold disks that no longer fulfill the Rayleigh-Jeans criteria, or when there is significant contamination from non-thermal sources such as stellar winds. Nonetheless, all of our Lupus sources in Figure~\ref{Fig7} are consistent with $\alpha_{\rm mm} > 2.0$ when considering uncertainties, as expected from grey body emission in the Rayleigh-Jeans regime. 

Figure~\ref{Fig7} also shows $\alpha_{\rm mm}$ as a function of $M_{\rm dust}$ (translated from $F_{\rm 1.33mm}$ using Equation~\ref{eqn-mass}). Contrary to previous studies of young disks that found no correlation between $\alpha_{\rm mm}$ and $M_{\rm dust}$ \cite[e.g.,][]{2005ApJ...631.1134A,2012A&A...540A...6R}, we find with our much more sensitive observations that low-mass disks follow an anti-correlation, followed by a flattening after $M_{\rm dust} \sim 5 M_{\oplus}$ as disks approach $\alpha_{\rm mm} \approx 2$. When considering only the full disks (see below), we fit the data with a piecewise linear relation:

{\footnotesize
\begin{equation} 
\alpha_{\rm mm} = \left\{
                              \begin{array}{ll}
                              -0.55 (\pm0.17) {\rm log} F_{\rm mm} + 2.62 (\pm0.09) \quad {\rm log} F_{\rm mm} \leq 0.81 \\ \\
                              +2.19(\pm0.05) \quad {\rm log} F_{\rm mm} > 0.81
                              \end{array}
                              \right.
\label{eqn-alpha}
\end{equation}
}

To test the significance of the anti-correlation, we apply a Spearman Rank test to the data where $F_{\rm 1.33mm} \leq 0.81$, which gives a probability of no correlation of $p = 0.005$. We note that using a simple linear relation also gives a statistically significant fit to the data, although an F-test could not identify which parametrization is more statistically significant. Regardless, this anti-correlation is not an observational bias: the gray regions in Figure~\ref{Fig7} show where we are not sensitive based on 3$\sigma$ limits, illustrating that our observational sensitivity does not influence the correlation. 

The decrease in $\alpha_{\rm mm}$ for brighter disks may be due to more efficient grain growth in higher-mass disks, which also tend to be around higher-mass stars (due to the $M_{\rm dust}$--$M_{\star}$ relation; Section~\ref{sec-dustmass}) and thus have faster dynamical timescales. If true, this rules out one of the scenarios proposed by \cite{2016ApJ...831..125P} to explain the steepening of the $M_{\rm dust}$--$M_{\star}$ relation with age, which is that grain growth is more efficient in disks around {\em lower}-mass stars. However, the decrease in $\alpha_{\rm mm}$ for higher-mass disks may also simply reflect larger optically thick regions, which would serve to artificially decrease $\alpha_{\rm mm}$ and thus mimic grain growth (e.g., \citealt{2017ApJ...845...44T}; van Terwisga et al., in prep). Higher resolution data that can resolve $\alpha_{\rm mm}$ as a function of radius are needed to distinguish between these scenarios. 

Figure~\ref{Fig7} shows that resolved transition disks tend to have higher $\alpha_{\rm mm}$ values when compared to the general protoplanetary disk population in Lupus. This is consistent with the findings of \cite{2014A&A...564A..51P}, who showed that $\alpha_{\rm mm}$ (between $\sim$1~mm and $\sim$3~mm; see their Table 2) is larger for transition disks compared to full protoplanetary disks in the Taurus, Ophiuchus, and Orion star-forming regions. They calculated a weighted mean and standard error on the mean of $\bar{\alpha}_{{\rm TD}} = 2.70 \pm 0.13$ and $\bar{\alpha}_{{\rm PPD}} = 2.20 \pm 0.07$. Here our wavelength range is smaller (890~$\mu$m--1.33~mm) but our disk population is from a single star-forming region and our observations are from uniform surveys at higher sensitivity. We find consistent results with $\bar{\alpha}_{{\rm TD}} = 2.70 \pm 0.10$ and $\bar{\alpha}_{{\rm PPD}} = 2.27 \pm 0.05$. \cite{2014A&A...564A..51P} explained the higher $\alpha_{\rm mm}$ values of transition disks in terms of the inner disk cavity. Namely, assuming transition disks had the same grain population as full disks before the inner disk clearing, and also that $\beta_{\rm d}$ increases with radius \cite[e.g.,][]{2011A&A...529A.105G, 2016A&A...588A..53T}, then disks with large inner cavities should lack large grains and therefore appear to have higher $\alpha_{\rm mm}$ values compared to full disks.

%===============================  RADIAL DRIFT ============================

\capstartfalse
\begin{figure*}
\begin{centering}
\includegraphics[width=17cm]{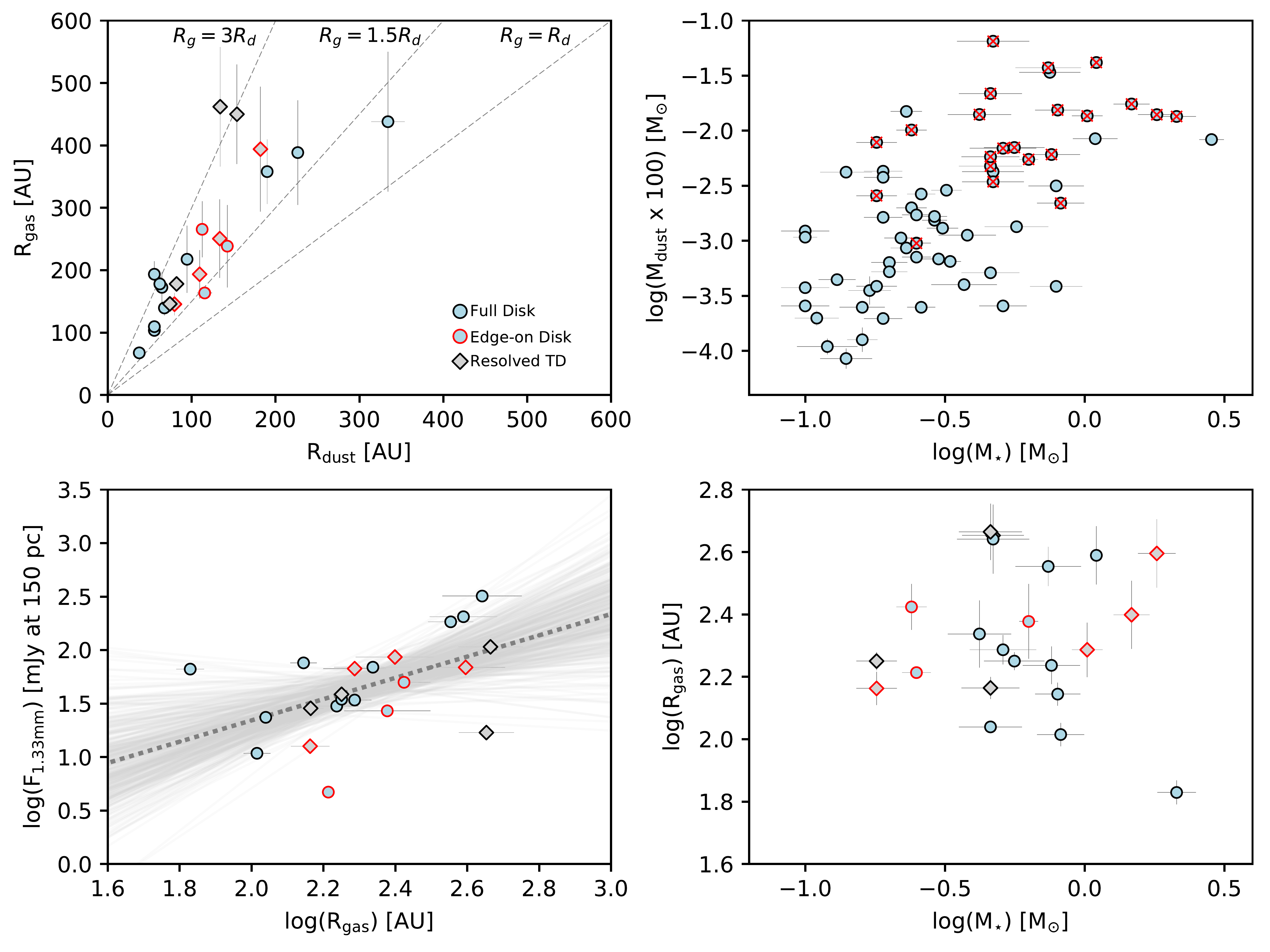}
\caption{\small { {\bf Top left:} Gas disk radii ($R_{\rm gas}$) compared to dust disk radii ($R_{\rm dust}$) for Lupus disks with constraints on both parameters: $R_{\rm gas}$ is universally larger than $R_{\rm dust}$ with an average ratio of $R_{\rm gas} / R_{\rm dust} = 1.96\pm0.04$ (Section~\ref{sec-radius}). Edge-on disks ($\mid i \mid \ge 65^{\circ}$) are outlined in red and transition disks (Section~\ref{sec-TD}) are indicated by diamonds. {\bf Top right:} the $M_{\rm dust}$--$M_{\star}$ correlation seen for Lupus disks (Paper~I); the sub-sample with $R_{\rm gas}$ measurements are highlighted by red crosses, illustrating the bias towards high-mass disks around high-mass stars. {\bf Bottom left:} a tentative correlation between $R_{\rm gas}$ and $F_{\rm 1.33~mm}$ (and thus disk mass), similar to the continuum size--luminosity relation seen previously for Lupus disks \citep{2017A&A...606A..88T}. The dashed gray line shows a Bayesian linear regression fit to the data and the light gray lines are a subsample of the MCMC chains. {\bf Bottom right:} the lack of correlation between $R_{\rm gas}$ and $M_{\star}$, likely due to the small range of $M_{\star}$ covered by the sub-sample of Lupus disks with $R_{\rm gas}$ constraints (see top right panel).}}
\label{Fig8}
\end{centering}
\end{figure*}
\capstartfalse

 \subsection{Dust Radial Drift? \label{sec-radii}}

Figure~\ref{Fig8} compares the gas radius ($R_{\rm gas}$) to the dust radius ($R_{\rm dust}$) for the 22 disks in our Lupus sample with constraints on both parameters, revealing that $R_{\rm gas}$ is universally larger than $R_{\rm dust}$. Although previous observations of individual disks have shown that the gas radius can extend beyond the dust radius, these were limited to the biggest and brightest disks \cite[e.g., TW Hya and HD~163296;][]{2009A&A...501..269P, 2012ApJ...744..162A, 2013A&A...557A.133D}. Here we show that this is a population-wide feature among young disks, and that $R_{\rm gas}$ is consistently $\approx1.5$--3.0$\times$$R_{\rm dust}$, with an average ratio of $R_{\rm gas} / R_{\rm dust} = 1.96\pm0.04$ (Section~\ref{sec-radius}).

The smaller dust disk sizes relative to the gas disks may be explained by dust grain growth and radial drift. Grain growth timescales are much shorter at smaller radial distances from the host star, resulting in grain size segregation with the larger grains preferentially located closer to the host star. In addition, as dust grains grow, drag forces from gas in sub-Keplerian rotation cause the larger solids to spiral inward toward the host star on short timescales \citep{1977Ap&SS..51..153W}. Both of these mechanisms can make the sub-mm/mm continuum emission appear smaller than the gas emission, because this continuum emission primarily traces larger (sub-mm/mm) grains while the gas emission primarily traces smaller (sub-$\mu$m) grains. Moreover, because dust growth and radial drift both produce similar particle size segregations in disks, it can be difficult to identify an unambiguous signature of radial drift (distinct from just grain growth) based on disk sizes alone.

However, numerical and analytical models also predict that a unique signature of radial drift is the {\em shape} of the sub-mm/mm continuum intensity profile  \cite[e.g.,][]{2014ApJ...780..153B}. This is because radial drift in the early phases of disk evolution, before grain growth in the outer disk has begun, naturally sets a distinct outer radius beyond which the disk is essentially devoid of dust; moreover, this sharp outer radius is preserved for sub-mm/mm grains in later phases of disk evolution, when grain growth and viscous spreading have set in. This reasoning has been used to invoke radial drift for explaining the differences in the sub-mm/mm dust and gas radii for TW Hya \citep{2012ApJ...744..162A, 2016A&A...586A..99H} and HD~163296 \citep{2013A&A...557A.133D}. The resolution of our data is insufficient to derive detailed continuum intensity profiles capable of fitting sharp outer edges, although higher-resolution ALMA observations of Lupus disks will be able to test for this signature of radial drift in the future. Nevertheless, radial drift is expected to come hand-in-hand with dust growth, as described above. 

An alternative explanation for the larger gas disk radii is that the $^{12}$CO emission is optically thick while the continuum emission is optically thin \citep{1998A&A...339..467G, 1998A&A...338L..63D}. Because the $^{12}$CO emission is optically thick, it is simply easier to detect small amounts of gas at large radii, whereas this is not the case for the optically thin emission from the dust. These optical depth effects can therefore produce similar observational signatures to dust grain growth and radial drift. Indeed, \cite{2017A&A...605A..16F} combined the dust evolution models from \cite{2015ApJ...813L..14B} with the thermo-chemical code DALI \citep{2012A&A...541A..91B, 2013A&A...559A..46B} to show that, at least for the case of the massive HD~163296 disk, the bulk of the difference between the gas and dust radii is due to the optical depth of the CO lines, with grain growth and radial drift having a more subtle effect on the steepness of the mm dust emission profile (see their Figures 16 and 17). 

\capstartfalse
\begin{figure*}[!ht]
\begin{centering}
\includegraphics[width=17cm]{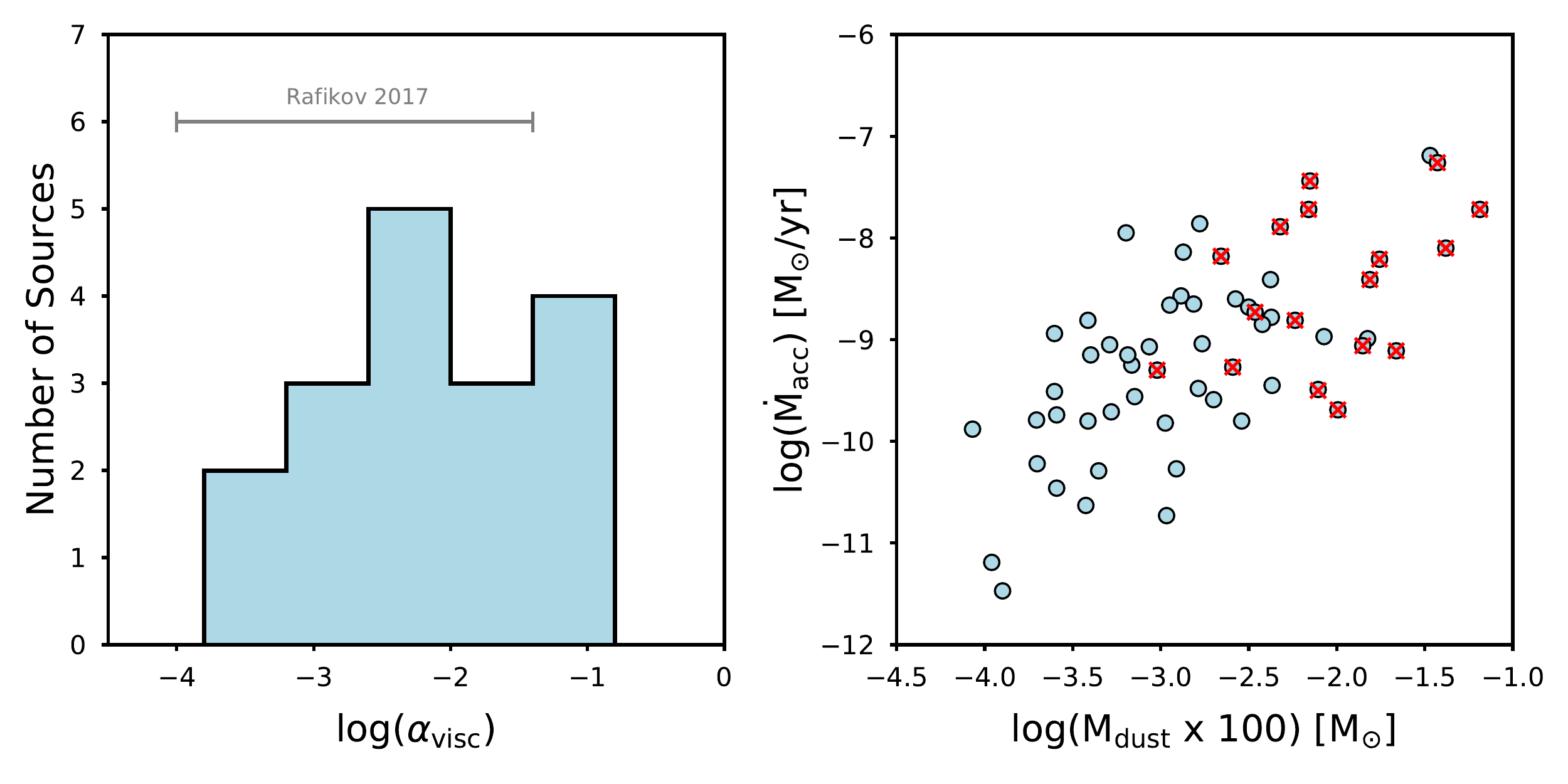}
\caption{\small { {\bf Left:} Distribution of the viscous parameter, $\alpha_{\rm visc}$, calculated from Equations~\ref{eqn-alpha} using outer disk radii derived from $^{12}$CO emission and disk masses derived from continuum emission (Section~\ref{sec-viscous}). The gray line shows the similarly large range of $\alpha_{\rm visc}$ values found by \cite{2017ApJ...837..163R}, who used continuum emission to derive both the disk radii and masses. {\bf Right:} the $\dot{M}_{\rm acc}$--$M_{\rm d}$ correlation seen for Lupus disks \citep{2016A&A...591L...3M}, with the sub-sample of resolved disks used for the $\alpha_{\rm visc}$ calculations highlighted by red crosses, illustrating the bias towards high-mass and strongly accreting disks.}}
\label{Fig9}
\end{centering}
\end{figure*}
\capstartfalse

To simulate more ``typical" Lupus disks, we update the \cite{2017A&A...605A..16F} models using $M_{\star} = 0.5$~$M_{\odot}$, $T_{\star} = 4700$~K, $\dot{M}_{\rm acc} = 10^{-9}$~$M_{\odot}$~yr$^{-1}$, $M_{\rm disk} = 10^{-4}$~$M_{\odot}$, and a tapered surface density profile with $\gamma = 1$ and $R_{\rm c} = 50$~au. The simulated images are then convolved with a $0\farcs25$ beam and use a distance of 150~pc. To test whether the disk size differences could be due solely to optical depth effects, we use a model with a uniform mix of small and large grains throughout the entire disk. Three other models then simulate grain growth, fragmentation, and radial drift by setting the maximum grain size as a function of radial distance from the host star based on fragmentation and radial drift limits with $\alpha_{\rm visc} = 10^{-2}, 10^{-3}, 10^{-4}$. 

Using the same curve-of-growth method described in Section~\ref{sec-radius} to measure disk radii, we find that $R_{\rm gas} / R_{\rm dust} \approx 1.5$ for the model with uniform grain sizes, and $R_{\rm gas} / R_{\rm dust} \approx 3.0$ for the models including grain growth and radial drift. Therefore, based on these models, optical depth effects could explain the lower range of the measured $R_{\rm gas} / R_{\rm dust}$ values in Lupus, but grain growth and radial drift may also be needed to explain the higher values of $R_{\rm gas} / R_{\rm dust}$ seen in the data. We caution that these models only consider the collisional growth and fragmentation of the dust grains and do {\em not} include their kinematics within the disk, which could increase the differences in the modeled gas and dust radii. Moreover, the models currently do not include simulated observational noise, which can affect the measured radii, especially for the gas due to low SNR in the outer disk. 

Additionally, Figure~\ref{Fig8} (lower left panel) shows a tentative correlation between $R_{\rm gas}$ and $F_{\rm 1.33~mm}$, analogous to the continuum size--luminosity relations seen previously in young disk populations \citep{2017A&A...606A..88T, 2017ApJ...845...44T}. The Bayesian linear regression method of \citep{2007ApJ...665.1489K} gives the correlation ${\rm log} F_{\rm 1.33mm} =  1.00 (\pm0.45) {\rm log} R_{\rm gas} - 0.66 (\pm1.04)$ with a correlation coefficient of $0.50\pm0.20$ and a dispersion of $0.42\pm0.08$. To test the significance of the correlation, we use a Spearman rank test, which gives $\rho=0.54$ and a p-value of 0.009. However, we caution that our sample is biased towards disks with both resolved continuum and gas emission; some Lupus disks exhibit faint and unresolved continuum emission, but bright and extended gas emission, thus may not follow this correlation.

We also do not see a correlation between $R_{\rm gas}$ and $M_{\star}$ (lower right panel of Figure~\ref{Fig8}), although this is likely due to the bias of our sample towards the highest-mass disks around the highest-mass stars (upper right panel of Figure~\ref{Fig8}). More sensitive and higher-resolution $^{12}$CO line observations can probe the gas disks around lower-mass stars to provide better constraints on these possible relations between fundamental disk parameters.

%=========================== VISCOUS EVOLUTION ============================

 \subsection{Viscous Disk Evolution \label{sec-viscous}}

Protoplanetary disks are traditionally thought to evolve through viscous evolution \citep{1974MNRAS.168..603L, 1998ApJ...495..385H}. According to viscous evolution theory, turbulence in the disk redistributes angular momentum by transporting it outward to larger radii over time, which in turn drives the accretion of disk material inward through the disk and onto the central star. The viscosity of the disk cannot be easily quantified, but it can be characterized by the so-called $\alpha$ prescription \citep{1973A&A....24..337S}, where:

\begin{equation}
\alpha_{\rm visc} =   \frac{ \dot{M}_{\rm acc}}{M_{\rm d}}  \frac{\mu}{k_{\rm B} T_{\rm d}}  \Omega r_{\rm d}^2.
\label{eqn-alpha}
\end{equation}

In this framework, $\alpha_{\rm visc}$ is a dimensionless parameter that is constant and $\lesssim1$, $T_{\rm d}$ is the disk temperature, $r_{\rm d}$ is the outer disk radius, $\mu$ is the mean molecular weight, $k_{\rm B}$ is the Boltzmann constant, and $\Omega$ is the Keplerian angular frequency where $\Omega = (GM_\star/r_{\rm d}^3)^{1/2}$. The ratio of the total disk mass ($M_{\rm d}$) to the stellar mass accretion rate ($\dot{M}_{\rm acc}$) gives the ``disk lifetime," which should be on the order of the age of the disk, but only for disk ages larger than the viscous timescale, $t_\nu$ \cite[e.g.,][]{2012MNRAS.419..925J, 2017MNRAS.468.1631R}

Placing observational constraints on $\alpha_{\rm visc}$ is important because this parameter is thought to be directly related to the angular momentum transport in disks, and thus critical for understanding disk evolution. However, obtaining these constraints is complicated by difficulties with measuring the masses and radii for large samples of disks. \cite{1998ApJ...495..385H} used the average observed properties of protoplanetary disks in Taurus and Chamaeleon~I, including a handful of disk sizes derived from 2.7~mm continuum observations, to estimate $\alpha_{\rm visc} \approx 10^{-2}$. However, recent observations of TW Hya and HD 163296, which provided the first tentative measurements of disk turbulence, suggest lower values of $\alpha_{\rm visc} \lesssim 10^{-3}$ \citep{2011ApJ...727...85H, 2015ApJ...813...99F, 2016A&A...592A..49T}.

For our Lupus sample, \cite{2017ApJ...837..163R} calculated $\alpha_{\rm visc}$ with Equation~\ref{eqn-alpha} using the $890~\mu$m continuum properties derived in Paper~I. He assumed $M_{\rm d} = 100~M_{\rm dust}$, took $r_{\rm d}$ from elliptical Gaussian fits to the continuum emission, and used isothermal disk temperatures of $T_{\rm d} = 20$~K. Additionally, he took $M_{\star}$ and $\dot{M}_{\rm acc}$ from \cite{2014A&A...561A...2A} and \cite{2017A&A...600A..20A}. \cite{2017ApJ...837..163R} found a wide range of $\alpha_{\rm visc}$ values spanning over two orders of magnitude, from $10^{-4}$ to 0.04, with no clustering around a particular value. The lack of any correlations between $\alpha_{\rm visc}$ and other global disk parameters ($M_{\rm d}$, $r_{\rm d}$, $\Sigma_{\rm d}$) or stellar parameters ($M_{\star}$, $R_{\star}$, $L_{\star}$) also lead him to suggest that angular momentum transport may actually be performed non-viscously, for example via MHD winds.

However, there are two major shortcomings of the analysis by \cite{2017ApJ...837..163R}: the small sample size of 26 sources (due to the need for well-resolved disks) and the use of Gaussian-fit estimates of the continuum emission for disk sizes (since the gas radii were not yet available). Our gas disk radii measured in Section~\ref{sec-radius} solve the latter issue, thus we repeat these $\alpha_{\rm visc}$ calculations using instead $r_{\rm d} = R_{\rm gas}$ in Equation~\ref{eqn-alpha}. As shown in the left panel of Figure~\ref{Fig9}, we still find a large range of $\alpha_{\rm visc}$ values spanning over two orders of magnitude, from $0.0003$ to $0.09$, which is not surprising given the tight relation between $R_{\rm gas}$ and  $R_{\rm dust}$ seen in Figure~\ref{Fig8}. 

Similar to \cite{2017ApJ...837..163R}, we also find no correlations between $\alpha_{\rm visc}$ and other disk or stellar properties. The reason for this is our small sample size of 22 disks, which does not allow us to overcome the first shortcoming of the work by \cite{2017ApJ...837..163R} described above. Importantly, these sub-samples of resolved Lupus disks are not only small, but also heavily biased towards the highest-mass disks, as shown in the right panel of Figure~\ref{Fig9}. This means that they lack sufficient leverage to exhibit the $\dot{M}_{\rm acc}$--$M_{\rm d}$ correlation, which is predicted by viscous evolution and seen in larger samples of disks in Lupus \citep{2016A&A...591L...3M} and the similarly young Chamaeleon~I region \citep{2017ApJ...847...31M}. The lack of correlations between $\alpha_{\rm visc}$ and other disk or stellar properties is thus dominated by the lack of correlation between $\dot{M}_{\rm acc}$ and $M_{\rm d}$ in the considered sub-samples. Interestingly, \cite{2017MNRAS.472.4700L} and \cite{2017ApJ...847...31M} have shown that the large scatter in the observed $\dot{M}_{\rm acc}$--$M_{\rm d}$ relation seen in Lupus and Chamaeleon~I can be reproduced by viscous evolution theory, if the age of the star-forming regions is smaller than the viscous timescales of the disks. If this is true, the assumption taken in Equation~\ref{eqn-alpha} of $\dot{M}_{\rm acc} / M_{\rm d} \gtrsim t_\nu$ no longer holds, and the derived values of $\alpha_{\rm visc}$ must be considered with great caution.

Ultimately, estimates of disk radii for systems with lower stellar and disk masses are needed to extend the sample to  the point where the $\dot{M}_{\rm acc}$--$M_{\rm d}$ relation can be recovered. Similar studies in older star-forming regions (e.g. Upper Sco) are also needed to insure that the $\dot{M}_{\rm acc} / M_{\rm d} \gtrsim t_\nu$ assumption is valid.

%=========================== SIZE LIMITS ============================

\subsection{Limits on the size of optically thick emission}
\label{sec-thick}

\begin{figure}
\includegraphics[width=7.6cm]{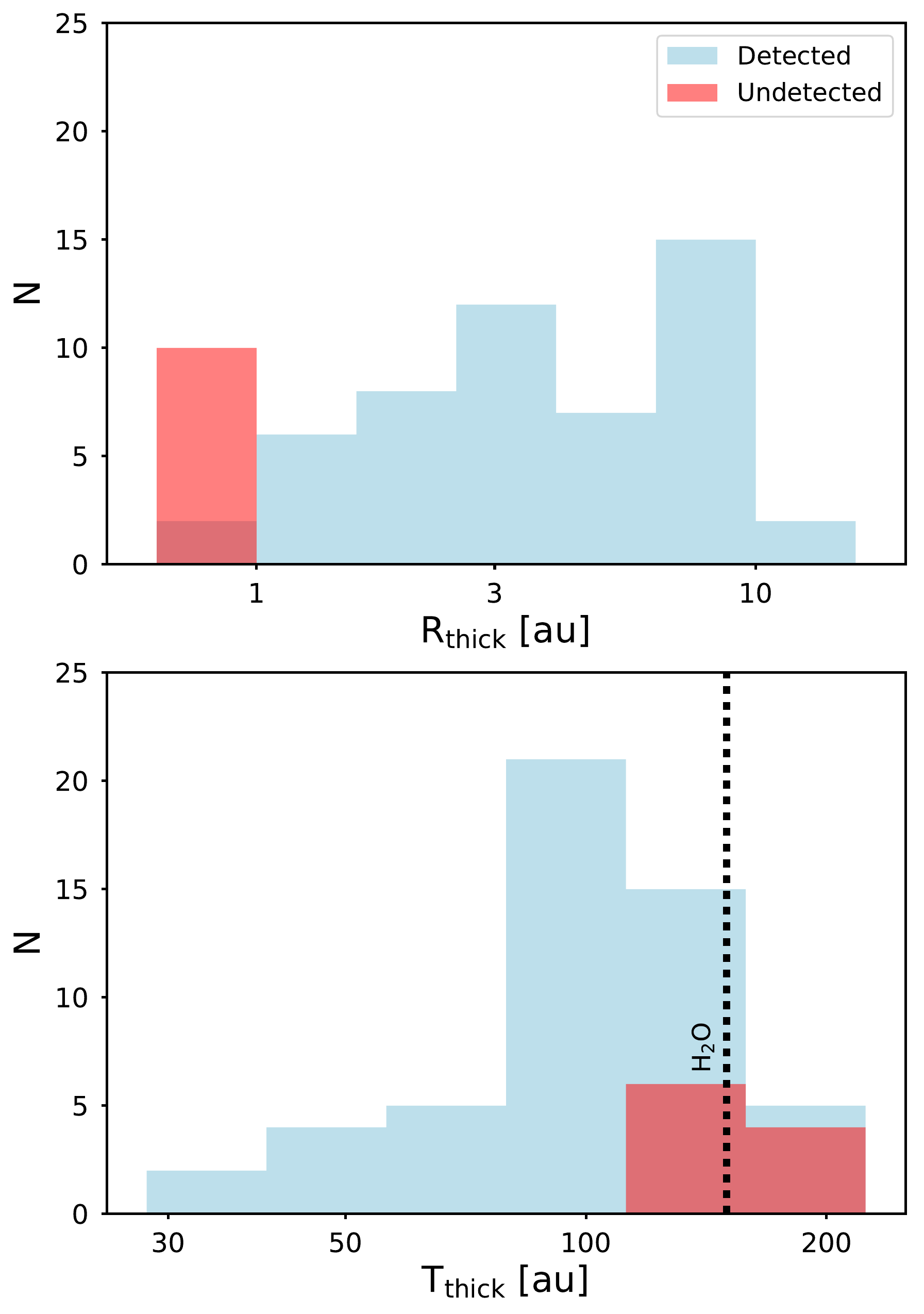}
\caption{Distributions of disk sizes (top) and outer disk temperatures (bottom) for Lupus disks, assuming optically thick mm emission. The blue and red histograms indicate our ALMA detections and non-detections, respectively (Section~\ref{sec-thick}). The dashed black line shows the water snowline.}
\label{Fig10}
\end{figure}

Sub-mm/mm continuum emission is generally optically thin in most regions of a protoplanetary disk, which is why it is the most commonly used tracer of dust mass (Section~\ref{sec-dustmass}). However, if the dust surface densities are high enough at the center of a disk---above roughly an Earth mass of dust spread over $\sim 10$\,au in diameter---the emission can become optically thick. In this case, the observed sub-mm/mm continuum flux is more a measure of the dust disk size, rather than the dust disk mass:

\begin{equation}
F_\nu = \int_{R_{\rm sub}}^{R_{\rm thick}} B_\nu(T)\,2\pi r\,dr \cos i / d^2,
\end{equation}

where $r$ is the radial distance in the disk, $d$ is the distance to the disk, $R_{\rm sub}$ is the inner radius of the disk defined by the dust sublimation temperature $T(R_{\rm sub}) = 1500$\,K, and $i$ is the inclination of the (geometrically thin) disk to the line-of-sight. The dust temperature is:

\begin{equation}
T = \left(\frac{L_{\star}}{16\pi\sigma_{\rm SB}r^2}\right)^{1/4},
\end{equation}

where $L_{\star}$ is the stellar luminosity and $\sigma_{\rm SB}$ is the Stefan-Boltzmann constant \cite[e.g.,][]{2010ARA&A..48..205D}.  For disks that are large enough to be resolved in our data, we derive $i$ from the aspect ratio of the disk; for the unresolved disks, we use a mean value of $\langle \cos\,i\rangle = 2/\pi$. We then integrate outwards until we match the observed central peak flux density in a beam centered on the stellar position, or for non-detections the corresponding $3\sigma$ upper limit. After removal of the transition disks \citep{2018arXiv180106154V} and unresolved binary system V856\,Sco, the derived $R_{\rm thick}$ ranges from 0.7 to 13~au with a median of 3.8~au for the detected disks, and 0.7 to 1.0~au with a median of 0.8~au for the non-detections (see upper panel of Figure~\ref{Fig10}).

For the non-detections, the limits on the dust content are very stringent, whether in terms of mass if the emission is optically thin or in radius if the emission is optically thick. We estimate that any optically thick central component would have to be less than 1\,au in radius for the disk to escape detection. For the detections, our analysis provides an upper limit to the size of any central optically thick component, as we are matching the central peak flux density in the $\sim$ 20~au beam and this likely includes some contribution from an extended optically thin component.

The derived limits on the thick disk radii of a few au are at an interesting scale, as this is comparable to where we expect the water snowline to reside. We expect the emission properties to change here due to the loss of ice and evolution of the grain size distribution \citep{2015ApJ...815L..15B}, as observed at larger radii in the much more luminous FU Orionis object V883 Ori \citep{2016Natur.535..258C}. We therefore re-plot these results in terms of the temperature of the outer edge of the optically thick disk (lower panel of Figure~\ref{Fig10}). The range for the detected disks is 31 to 193\,K with a median of 104\,K and 123 to 223~K with a median of 104\,K for the non-detections.

If a substantial fraction of the observed emission does indeed come from a compact and optically thick region, then many of the Lupus disks should remain detectable in relatively short ALMA integrations in much more extended configurations. Small and bright optically thick cores are found in the ultra-high resolution ALMA observations of TW Hya \citep{2016ApJ...820L..40A} and HL Tau \citep{2015ApJ...808L...3A}. If future imaging surveys show that such features are common, it would indicate the existence of large reservoirs of material in the innermost regions of disks as required by models of {\it in-situ} formation for short-period planets  \citep{2013MNRAS.431.3444C}.

%===============================  SUMMARY ============================

\section{Summary}
 \label{sec-summary}
 
We have conducted a high-sensitivity, high-resolution survey in ALMA Band~6 (1.33~mm) of a complete sample of protoplanetary disks in the young (1--3~Myr) and nearby (150--200~pc) Lupus star-forming region. The proximity and youth of this region make it an ideal target for a baseline study of early disk properties. This work built off of Paper~I, which used ALMA Band~7 (890~$\mu$m) observations to constrain both dust and gas masses for an unbiased sample of protoplanetary disks in Lupus. In this work, we expanded our statistical studies by using our Band~6 data to estimate gas disk sizes and, by combining these new observations with our previous Band~7 data, constrained disk evolution processes such as dust grain growth, radial drift, and viscous evolution.

\begin{itemize}

\item Our complete sample contained 95 Lupus protoplanetary disks, for which we obtained ALMA Band~6 data in the 1.33~mm continuum and $^{12}$CO, $^{13}$CO, and C$^{18}$O $J=2$--1 lines. We detected 71 disks in the continuum, 48 in $^{12}$CO, 20 $^{13}$CO, and 8 in C$^{18}$O. The typical spatial resolution of our observations was $0\farcs25$ with a medium 3$\sigma$ continuum sensitivity of 0.30~mJy.

\item We used the continuum and $^{12}$CO emission to estimate the dust and gas radii for 22 Lupus disks. We employed a ``Keplerian masking" technique to enhance the SNR of the $^{12}$CO emission in the outer disk regions. We found that $R_{\rm gas}$ is universally larger than $R_{\rm dust}$, with an average ratio of $R_{\rm gas} / R_{\rm dust} = 1.96\pm0.04$. This is likely due to both the optically thick $^{12}$CO emission as well as the growth and inward drift of the dust. We also found a tentative correlation between $R_{\rm gas}$ and $F_{\rm 1.33 mm}$, reminiscent of the continuum size-luminosity relation seen in young star-forming regions.

\item Similar to Paper~1, we used the continuum emission to constrain disk dust masses down to $\sim$$0.2~M_{\oplus}$. We recovered the $M_{\rm dust}$--$M_{\star}$ relation and used a stacking analysis to again show that the average dust mass of an undetected Lupus disk is comparable to debris disk levels, indicating that protoplanetary disks evolve rapidly once clearing begins.

\item We combined our Band~6 and Band~7 data to measure the mm spectral index, $\alpha_{\rm mm}$, for 70 Lupus disks down to $F_{\rm 1.33mm} =0.35$~mJy. We found an anti-correlation between $\alpha_{\rm mm}$ and $M_{\rm dust}$ for low-mass disks ($M_{\rm dust} \lesssim 5 M_{\oplus}$), followed by a flattening to $\alpha_{\rm mm} \approx 2$. The decrease in $\alpha_{\rm mm}$ for brighter disks may be due to more efficient grain growth in higher-mass disks, or may simply reflect larger optically thick regions in more massive disks, although our current data cannot distinguish between these scenarios. 

\item Using our $R_{\rm gas}$ measurements, we calculated the viscous parameter, $\alpha_{\rm visc}$, finding a large range of values spanning several orders of magnitude and no correlations with other disk or stellar properties. We attributed this to the small and biased sample, which is too limited to recover the $\dot{M}_{\rm acc}$--$M_{\rm d}$ relation seen in larger samples of Lupus disks. Estimates of disk radii for systems with lower stellar and disk masses are thus still needed.

\item We placed constraints on the sizes of optically thick inner disk regions for both the continuum detections and non-detections in our sample. The derived limits of a few au are interesting because they are comparable to the expected location of the water snowline, where sub-mm/mm emission properties should change. If a substantial fraction of the observed continuum emission does indeed come from compact and optically thick inner disk regions, then this could potentially provide a large reservoir of material for the {\it in-situ} formation of short-period planets.

\end{itemize}
 
\begin{acknowledgements}

MA and JPW were supported by NSF and NASA grants AST-1208911 and NNX15AC92G, respectively. MA also acknowledges support from NSF AST-1518332, NASA NNX15AC89G and NNX15AD95G/NEXSS, and the Center for Integrative Planetary Science. Leiden is supported by the European Union A-ERC grant 291141 CHEMPLAN, by the Netherlands Research School for Astronomy (NOVA), and by grant 614.001.352 from the Netherlands Organization for Scientific Research (NWO). AM and CFM are supported by ESO Fellowships. MRH is supported by NWO grant 614.001.352. MT has been supported by the DISCSIM project, grant agreement 341137 funded by the European Research Council under ERC-2013-ADG. This paper makes use of the following ALMA data: 2015.1.00222.S, 2016.1.01239.S, 2013.1.00226.S, 2013.1.01020.S. ALMA is a partnership of ESO (representing its member states), NSF (USA) and NINS (Japan), together with NRC (Canada), NSC and ASIAA (Taiwan), and KASI (Republic of Korea), in cooperation with the Republic of Chile. The Joint ALMA Observatory is operated by ESO, AUI/NRAO and NAOJ. The National Radio Astronomy Observatory is a facility of the National Science Foundation operated under cooperative agreement by Associated Universities, Inc. This work benefited from NASA's Nexus for Exoplanet System Science (NExSS) research coordination network sponsored by NASA's Science Mission Directorate.

\end{acknowledgements}

%===============================  REFERENCES ============================

% \bibliography{../bib.bib}

%===============================  APPENDIX ============================

\newpage

\appendix

\setcounter{figure}{0}
\renewcommand{\thefigure}{A\arabic{figure}}

\section{A. $^{12}$CO Spectra}
\label{app-co}

\capstartfalse
\begin{figure*}[!ht]
\begin{centering}
\includegraphics[width=15cm]{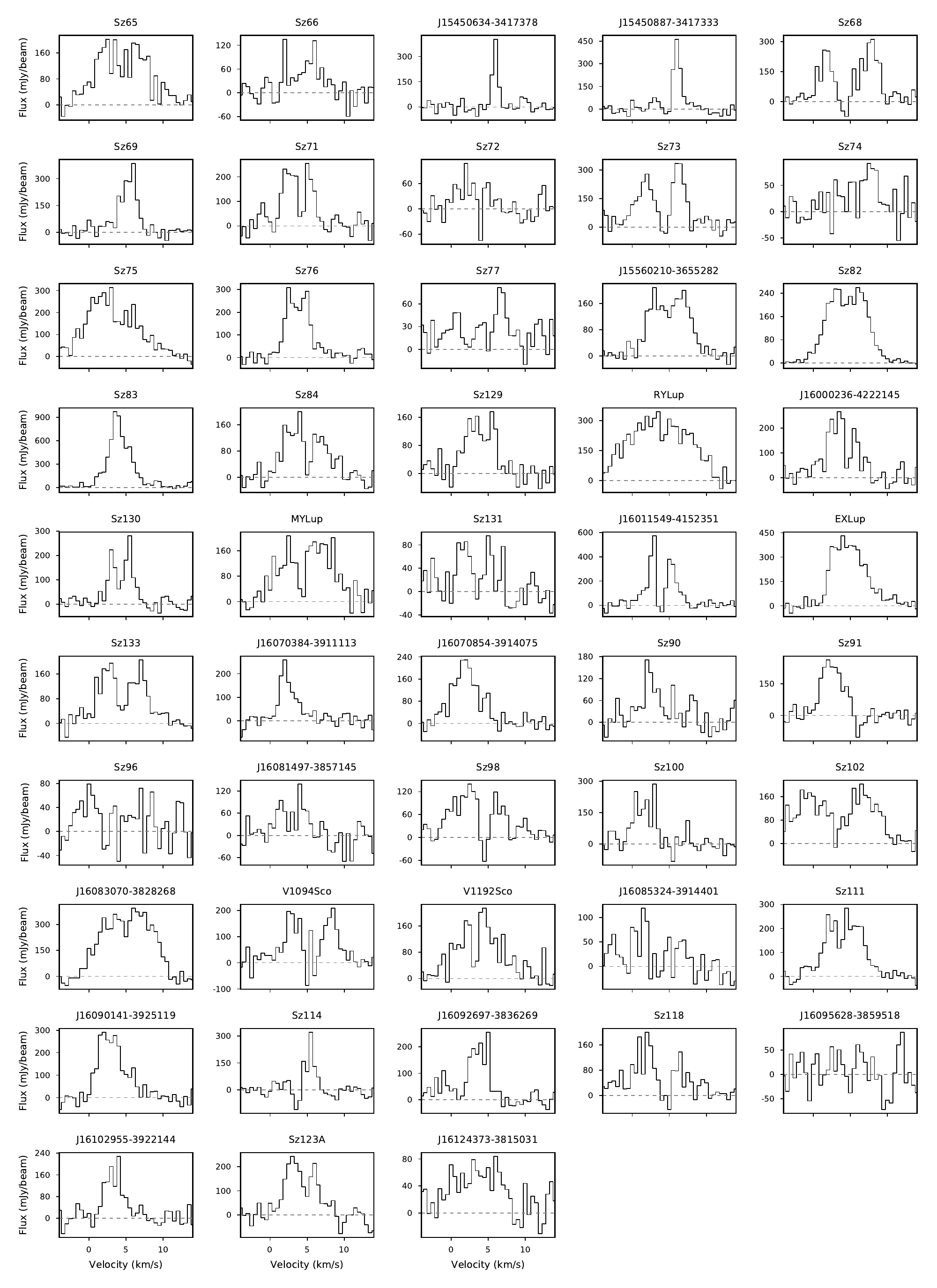}
\caption{\small {$^{12}$CO spectra for the 48 Lupus disks detected in this line, illustrating that cloud absorption is common.}}
\label{FigA1}
\end{centering}
\end{figure*}
\capstartfalse

\clearpage

\setcounter{figure}{0}
\renewcommand{\thefigure}{B\arabic{figure}}

\section{B. Keplerian Masking}
\label{app-radii}

\begin{figure}[!ht]
\begin{centering}
\includegraphics[width=8cm]{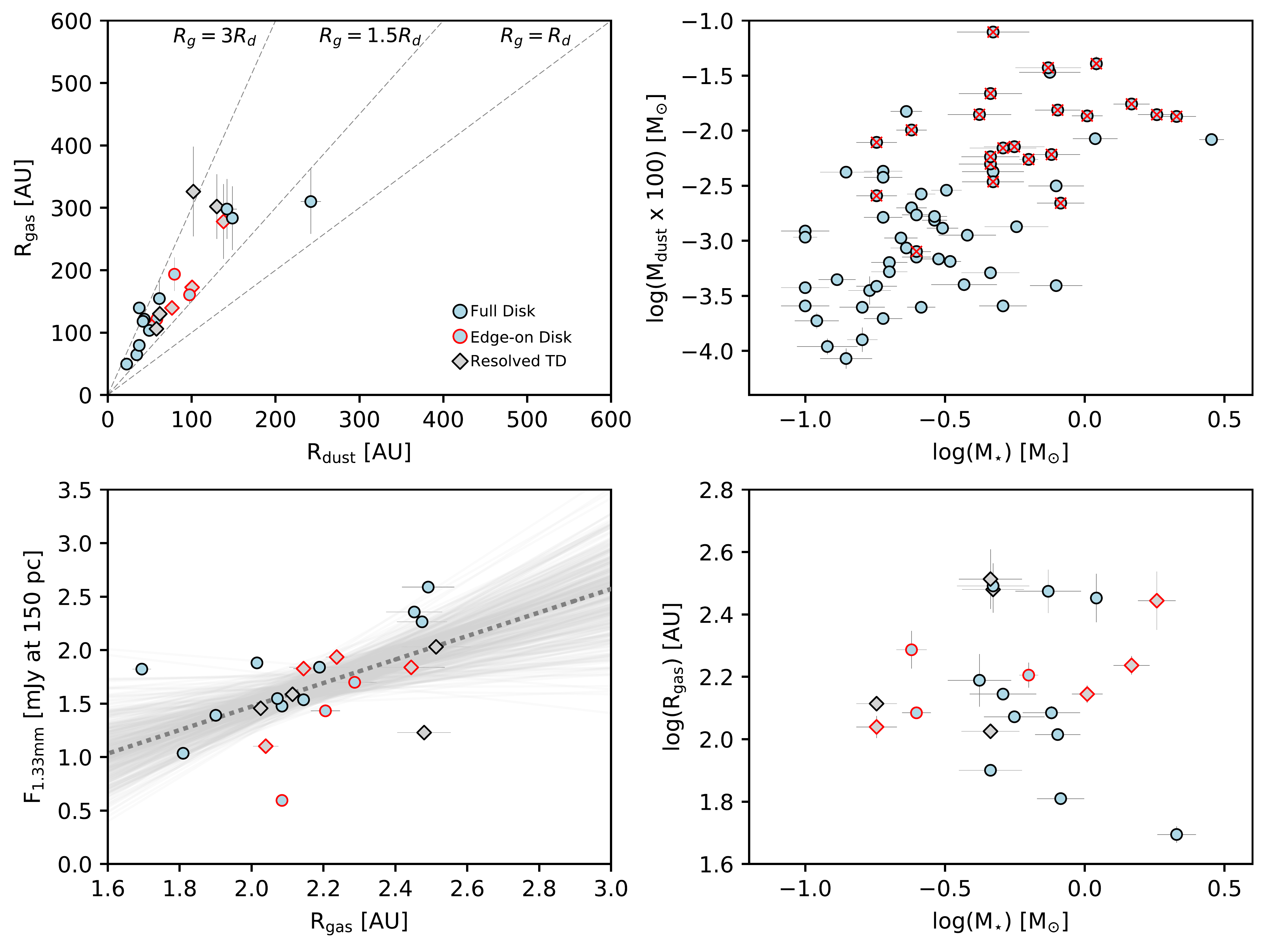}
\caption{\small {Comparisons of the gas disk radius ($R_{\rm gas}$) and dust disk radius ($R_{\rm dust}$) for Lupus disks with constraints on both parameters, now using 68\% of the total flux for the radius measurements (rather than 90\% as in Section~\ref{sec-radius}). The plot is remarkably similar to Figure~\ref{Fig8}: $R_{\rm gas}$ is still universally larger than $R_{\rm dust}$ with a similar average ratio of $R_{\rm gas} / R_{\rm dust} = 2.06\pm0.03$.}}
\label{FigB1}
\end{centering}
\end{figure}

Figure~\ref{FigB1} shows the same as the upper left panel of Figure~\ref{Fig8}, except now using 68\% of the total flux as the cutoff for the radius curve-of-growth measurements (rather than 90\%; see Section~\ref{sec-radius}). Significant differences in radius estimates can arise when using different flux cutoffs in a curve-of-growth analysis due to low SNRs in the outer disk regions where emission can be very faint, especially for the gas. However, as shown in Figure~\ref{FigB1}, using a more conservative flux cutoff does not change our key results: we still find universally larger gas disk sizes with a similar average ratio of $R_{\rm gas} / R_{\rm dust} = 2.06\pm0.03$.

\capstartfalse
\capstartfalse

\begin{figure}[!ht]
\begin{centering}
\includegraphics[width=8cm]{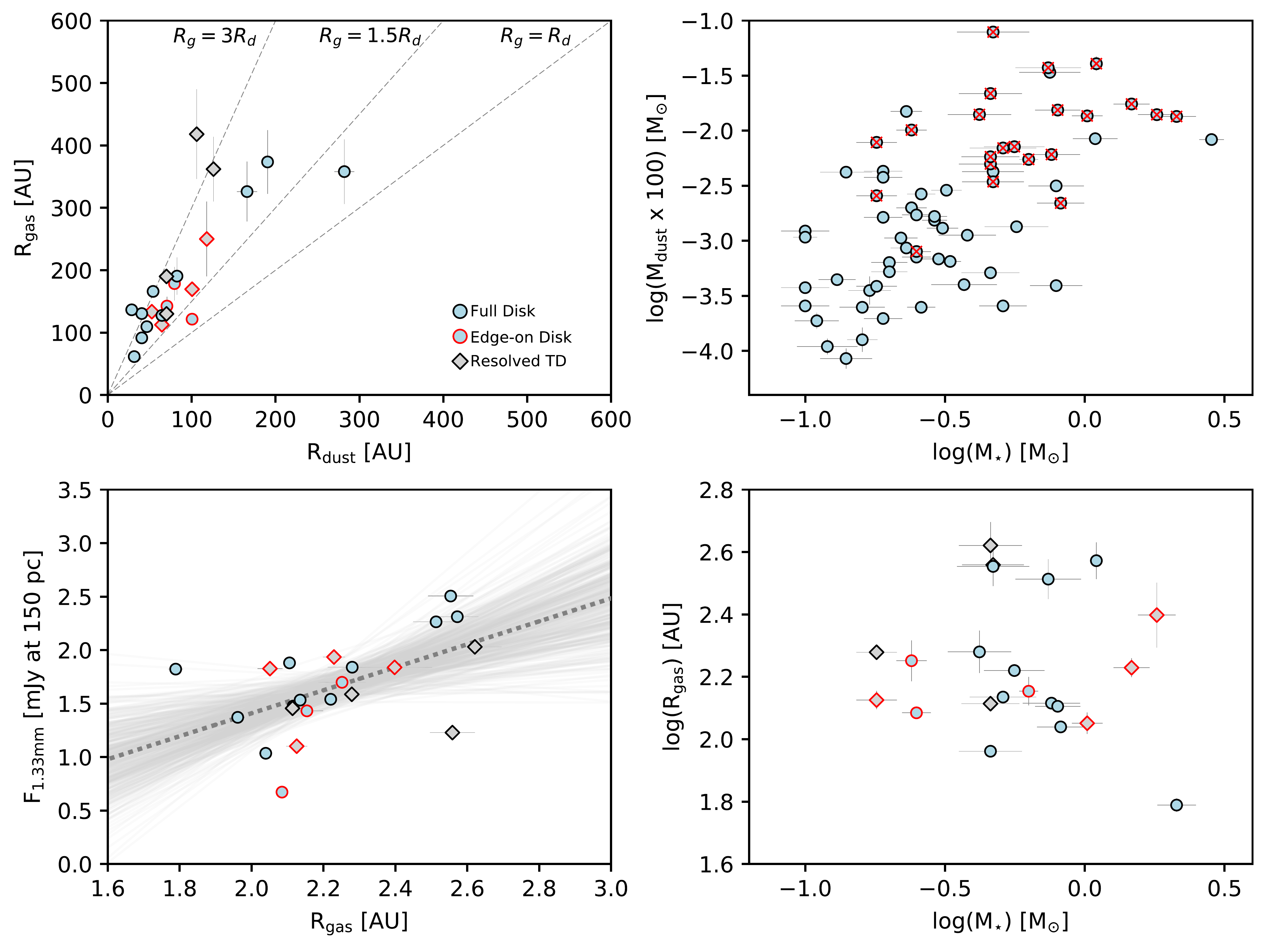}
\caption{\small {Comparisons of the gas disk radius ($R_{\rm gas}$) and dust disk radius ($R_{\rm dust}$) for Lupus disks with constraints on both parameters, now using a circular aperture (rather than an elliptical aperture) to measure the radius at 90\% of the total flux. The plot is remarkably similar to Figure~\ref{Fig8}: $R_{\rm gas}$ is still universally larger than $R_{\rm dust}$ with a similar average ratio of $R_{\rm gas} / R_{\rm dust} = 1.90\pm0.03$.}}
\label{FigB2}
\end{centering}
\end{figure}

Figure~\ref{FigB2} shows the same as the upper left panel of Figure~\ref{Fig8}, except now using a circular aperture (rather than an elliptical aperture) to measure the radius at 90\% of the total flux. Significant variations in radius estimates can arise when using elliptical apertures, in particular for edge-on disks, if the inclination is poorly constrained. However, as shown in Figure~\ref{FigB2}, using a circular aperture does not change our key results: we still find universally larger gas disk sizes with a similar average ratio of $R_{\rm gas} / R_{\rm dust} = 1.90\pm0.03$. 

\capstartfalse
\capstartfalse

\begin{figure*}
\begin{centering}
\includegraphics[width=17.8cm]{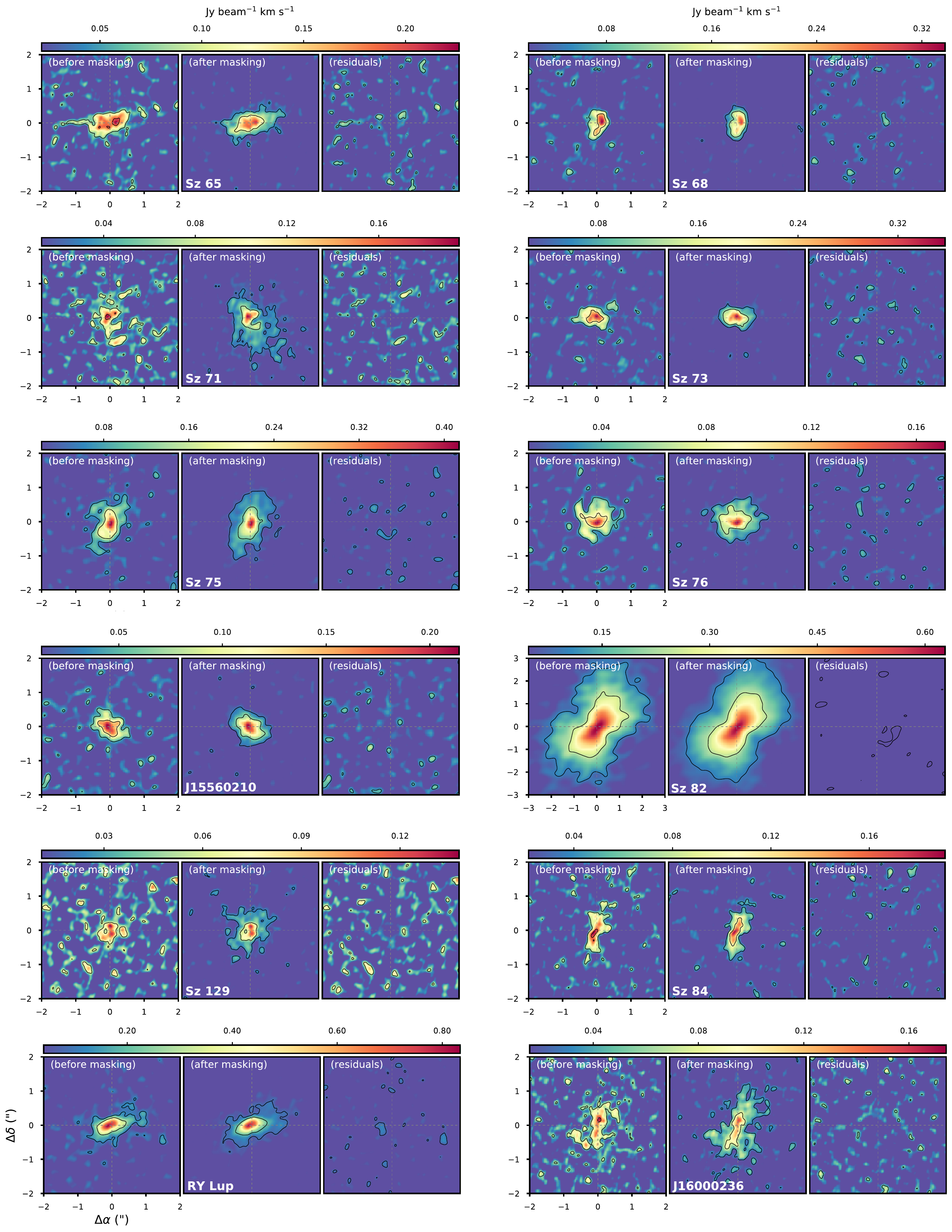}
\caption{\small {Zero-moment maps for our Lupus sources with $R_{\rm gas}$ measurements, before (left) and after (middle) Keplerian masking, as well as the residuals (right). The black lines are 2$\sigma$ and 5$\sigma$ contours, illustrating the improvement in SNR in the fainter outer disk regions (see Section~\ref{sec-radius}). The residuals confirm that the masking is not excluding any disk flux.}}
\label{FigB3}
\end{centering}
\end{figure*}

Figure~\ref{FigB3} shows the zero-moment maps before/after Keplerian masking for all Lupus disk with measured gas radii.

\capstartfalse
\capstartfalse

\renewcommand{\thefigure}{B\arabic{figure} (Cont.)}
\addtocounter{figure}{-1}

\capstartfalse
\begin{figure*}
\begin{centering}
\includegraphics[width=17.8cm]{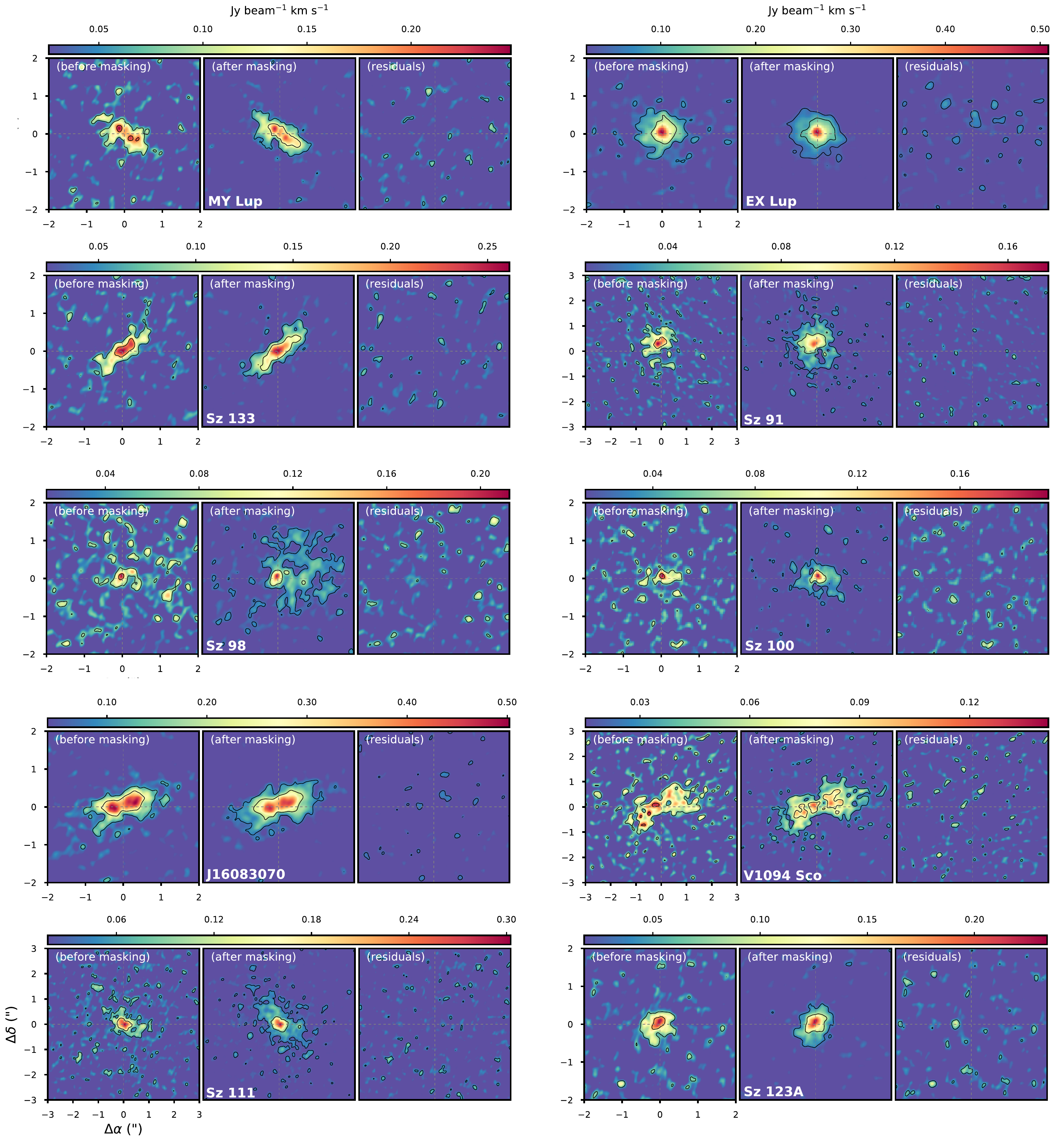}
\caption{}
\label{FigB3b}
\end{centering}
\end{figure*}
\capstartfalse

\renewcommand{\thefigure}{\arabic{figure}}

\clearpage

\setcounter{figure}{0}
\renewcommand{\thefigure}{C\arabic{figure}}
\section{C. Possible Outflow Sources}
\label{app-outflow}

\capstartfalse
\begin{figure*}[!ht]
\begin{centering}
\includegraphics[width=18cm]{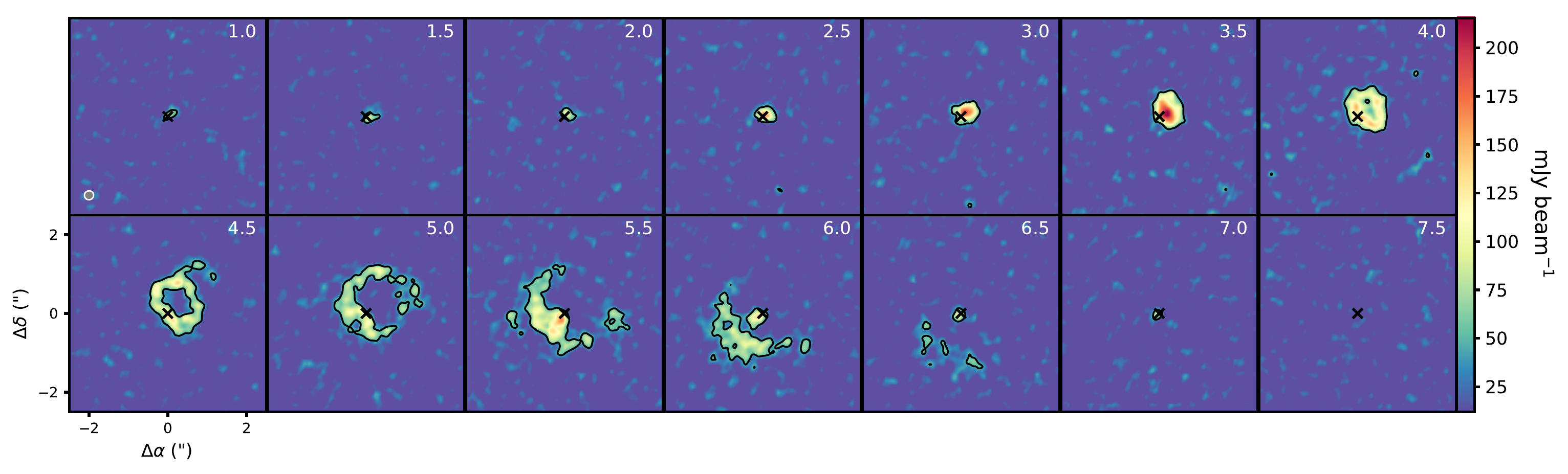}
\caption{\small {Possible outflow seen in the $^{12}$CO channel maps of Sz~83. The black cross is the location of the continuum source and the black lines are 4$\sigma$ contours of the $^{12}$CO emission. The velocities in km~s$^{-1}$ are given in the top right corner of each channel and the beam size is shown by the gray ellipse in the lower left corner of the first channel.}}
\label{FigC1}
\end{centering}
\end{figure*}
\capstartfalse

\capstartfalse
\begin{figure*}[!ht]
\begin{centering}
\includegraphics[width=18cm]{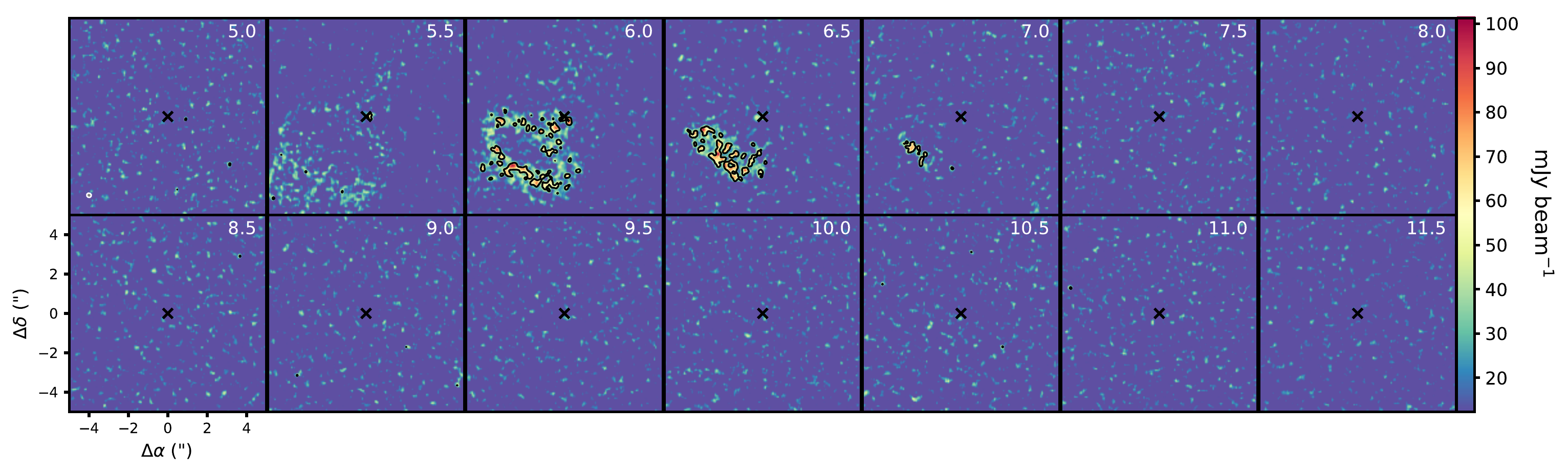}
\caption{\small {Possible outflow seen in the $^{12}$CO channel maps of J15450634-3417378. The symbols are the same as in Figure~\ref{FigC1}.}}
\label{FigC2}
\end{centering}
\end{figure*}
\capstartfalse

\capstartfalse
\begin{figure*}[!ht]
\begin{centering}
\includegraphics[width=18cm]{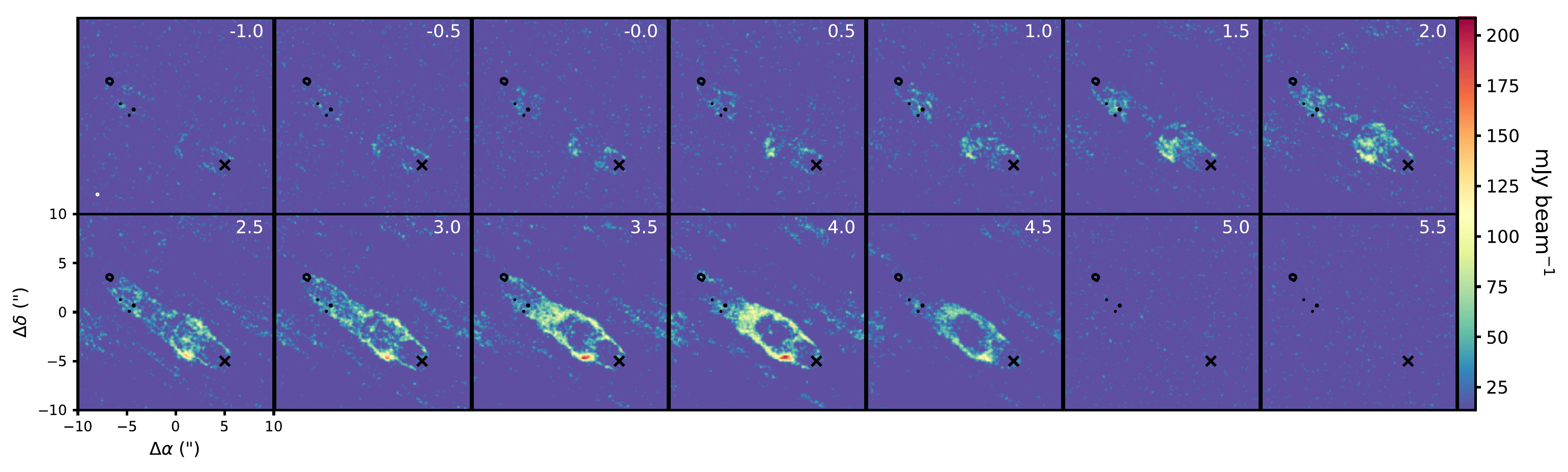}
\caption{\small {Outflow from IRAS~15398-3359, seen in the $^{12}$CO channel maps of J15430131-3409153. The symbols are the same as in Figure~\ref{FigC1}, except the contours now trace the continuum emission in order to highlight the location of IRAS~15398-3359 (upper left corner). Moreover, because we do not detect J15430131-3409153, the black cross is the phase center of our observations.}}
\label{FigC3}
\end{centering}
\end{figure*}
\capstartfalse

\end{document}